\documentclass[10pt,journal,compsoc]{IEEEtran}

\usepackage{amsmath,amssymb,amsfonts}
\usepackage{graphicx}
\usepackage{textcomp}
\usepackage{xcolor}
\usepackage [english]{babel}
\usepackage [autostyle, english = american]{csquotes}
\usepackage{amsmath}
\usepackage{graphicx}
 \usepackage{enumitem}
\usepackage{float}
\usepackage{graphics} % for pdf, bitmapped graphics files
\usepackage{epsfig} 
\usepackage{tabularx}
\usepackage{multicol}
\usepackage{multirow}
\usepackage{tabularx}
\usepackage{pdflscape}
\usepackage{bbding}
\usepackage{wasysym}
\usepackage{amssymb}% http://ctan.org/pkg/amssymb
\usepackage{pifont}% http://ctan.org/pkg/pifont
\usepackage{amssymb}
\usepackage{booktabs}
\usepackage{afterpage}
\usepackage{bbm}
\usepackage{subfigure}
\usepackage{capt-of}
\usepackage{rotating}
\usepackage{lscape}
\usepackage{hyperref}
\hypersetup{bookmarks=false}
\usepackage{pgfplots}
\usepackage{wrapfig}
\pgfplotsset{width=8cm,compat=1.9}
\graphicspath{ {images/} }
\usepackage{amsmath}
\usepackage{braket} % needed for \Set
\usepackage{caption}
\usepackage{algorithm}
\usepackage{xcolor}
\usepackage{xr}
\usepackage[noend]{algpseudocode}
\usepackage{mathtools,bm}

\DeclareMathOperator*{\argmax}{arg\,max}

\newcommand{\blue}[1]{\textcolor{black}{#1}}

\makeatletter
\newcommand*{\addFileDependency}[1]{% argument=file name and extension
  \typeout{(#1)}
  \@addtofilelist{#1}
  \IfFileExists{#1}{}{\typeout{No file #1.}}
}
\makeatother
\newcommand*{\myexternaldocument}[1]{%
    \externaldocument{#1}%
    \addFileDependency{#1.tex}%
    \addFileDependency{#1.aux}%
}

\myexternaldocument{supp}

\begin{document}
\bstctlcite{IEEEexample:BSTcontrol}

\title{SplitPlace: AI Augmented Splitting and Placement of Large-Scale Neural Networks in Mobile Edge Environments\thanks{A preliminary version of this work was presented at the Student Research Competition in ACM SIGMETRICS Conference 2021~\cite{tuli2021splitplace}.}}

%%%%%%%%%%%%%%%%%%%%%%%%%%%%Authors%%%%%%%%%%%%%%%%%%%%%%%%%%%%%%%%%%%%%%%%%%
\author{
        Shreshth~Tuli,
        Giuliano~Casale
    and~Nicholas~R.~Jennings% <-this % stops a space
\IEEEcompsocitemizethanks{
\IEEEcompsocthanksitem S. Tuli, G. Casale and N. R. Jennings are with the Department
of Computing, Imperial College London, United Kingdom. 
\IEEEcompsocthanksitem N. R. Jennings is also with Loughborough University, United Kingdom.\protect \\E-mails: \{s.tuli20, g.casale\}@imperial.ac.uk, n.r.jennings@lboro.ac.uk.\protect
}% <-this % stops an unwanted space
\thanks{Manuscript received ---; revised ---.}}

%%%%%%%%%%%%%%%%%%%%%%%%%%%%End-Authors%%%%%%%%%%%%%%%%%%%%%%%%%%%%%%%%%%%%%%%%%%

\markboth{IEEE Transactions on Mobile Computing}%
{Tuli \MakeLowercase{\textit{et al.}}: --- }

\IEEEtitleabstractindextext{%
\begin{abstract}
In recent years, deep learning models have become ubiquitous in industry and academia alike. Deep neural networks can solve some of the most complex pattern-recognition problems today, but come with the price of massive compute and memory requirements. This makes the problem of deploying such large-scale neural networks challenging in resource-constrained mobile edge computing platforms, specifically in mission-critical domains like surveillance and healthcare. To solve this, a promising solution is to split resource-hungry neural networks into lightweight disjoint smaller components for pipelined distributed processing. At present, there are two main approaches to do this: semantic and layer-wise splitting. The former partitions a neural network into parallel disjoint models that produce a part of the result, whereas the latter partitions into sequential models that produce intermediate results. However, there is no intelligent algorithm that decides which splitting strategy to use and places such modular splits to edge nodes for optimal performance. To combat this, this work proposes a novel AI-driven online policy, SplitPlace, that uses Multi-Armed-Bandits to intelligently decide between layer and semantic splitting strategies based on the input task's service deadline demands. SplitPlace places such neural network split fragments on mobile edge devices using decision-aware reinforcement learning for efficient and scalable computing. Moreover, SplitPlace fine-tunes its placement engine to adapt to volatile environments. Our experiments on physical mobile-edge environments with real-world workloads show that SplitPlace can significantly improve the state-of-the-art in terms of average response time, deadline violation rate, inference accuracy, and total reward by up to 46, 69, 3 and 12 percent respectively.
\end{abstract}

}

\urlstyle{tt}
\maketitle

\IEEEdisplaynontitleabstractindextext

\IEEEpeerreviewmaketitle

\IEEEraisesectionheading{\section{Introduction}\label{sec:introduction}}
% the growth of neural networks
% modern networks cpu and ram requirements very high
Modern Deep Neural Networks (DNN) are becoming the backbone of many industrial tasks and activities~\cite{gill2019transformative}.  As the computational capabilities of devices have improved, new deep learning models have been proposed to provide improved performance~\cite{zhu2018benchmarking, li2019edge}. Moreover, many recent DNN models have been incorporated with mobile edge computing to give low latency services with improved accuracies compared to shallow networks, particularly in complex tasks like image segmentation, high frame-rate gaming and traffic surveillance~\cite{khanna2020intelligent}. The performance of such neural models reflects directly on the reliability of application domains like self-driving cars, healthcare and manufacturing~\cite{kraemer2017fog, gill2019transformative}. However, to provide high accuracy, such neural models are becoming increasingly demanding in terms of data and compute power, resulting in many challenging problems.  To accommodate these increasing demands, such massive models are often hosted as web services deployed on the public cloud~\cite{zhang2017deep, roopaei2017deep}.

% Challenge of response times
\textbf{Challenges.} Recently, application demands have shifted from either high-accuracy or low-latency to both of these together, termed as HALL (high-accuracy and low-latency) service delivery~\cite{gill2019transformative}. Given the prevalence and demand of DNN inference, serving them on a public cloud with tight bounds of latency, throughput and cost is becoming increasingly challenging~\cite{gunasekaran2020implications}. In this regard, recent paradigms like mobile edge computing seem promising. Such approaches allow a robust and low-latency deployment of Internet of Things (IoT) applications close to the edge of the network. Specifically, to solve the problem of providing HALL services, recent work proposes to integrate large-scale deep learning models with modern frameworks like edge computing~\cite{gunasekaran2020implications, laskaridis2020spinn, liang2020ai}. However, even the most recent approaches either provide a low Service Level Agreement (SLA) violation mode or a high-accuracy mode~\cite{laskaridis2020spinn, gunasekaran2020implications} and struggle to provide the benefits of both modes at the same time. 

% Challenge of resource limitation
Another challenge of using edge computing is that mobile edge devices face severe limitations in terms of computational and memory resources as they rely on low power energy sources like batteries, solar or other energy scavenging methods~\cite{abbas2017mobile, mao2016dynamic}. This is not only because of the requirement of low cost, but also the need for mobility in such nodes~\cite{khanna2020intelligent}. In such systems, it is possible to handle the processing limitations of massive DNN models by effective preemption and prolonged job execution. However, memory bottlenecks are much harder to solve~\cite{shao2020communication}. In a distributed edge environment, storage spaces are typically mapped to network-attached-storage (NAS) media. Thus, prior work that runs inference on a pre-trained DNN without memory-aware optimizations leads to high network bandwidth overheads due to frequent overflow of memory and the use of virtual memory (swap space) on NAS media, making high fidelity inference using DNNs hard~\cite{liu2020scale, laskaridis2020spinn}. To deploy an upgraded AI model, tech giants like Amazon, Netflix and Google usually consider completely revamping their infrastructures and upgrading their devices, raising many sustainability concerns~\cite{gill2019transformative}. This has made the integration of massive neural network models with such devices a challenging and expensive ordeal. 

\begin{figure}[t]
    \centering \setlength{\belowcaptionskip}{-5pt}
    \includegraphics[width=\linewidth]{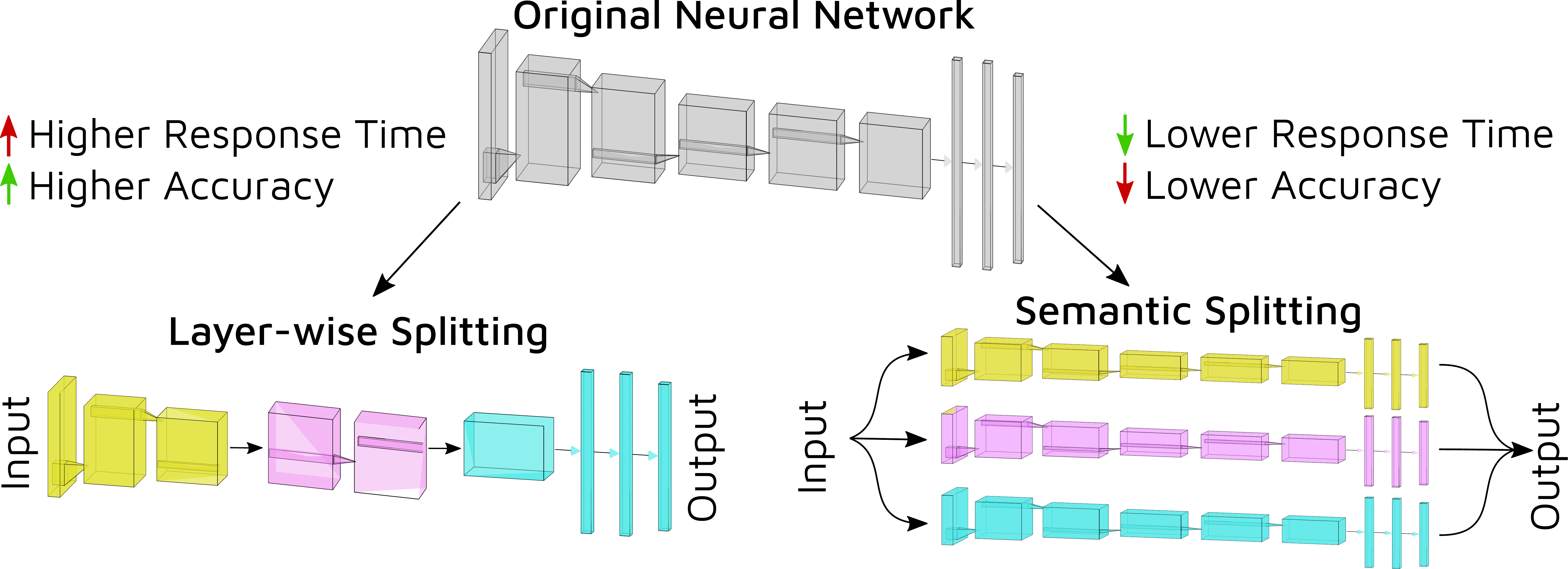}
    \caption{Overview of layer and semantic splitting strategies}
    \label{fig:scale}
\end{figure} 

% Ray of hope -> large DNNs on legacy infra
\textbf{Solution.} A promising solution for this problem is the development of strategies that can accommodate large-scale DNNs within legacy infrastructures. However, many prior efforts in this regard~\cite{kim2017splitnet, shi2020communication, lim2020federated} have not yet tackled the challenge of providing a holistic strategy for not only distributed learning, but also inference in such memory-constrained environments. Recently, research ideas have been proposed like Cloud-AI, Edge-AI and Federated learning that aim to solve the problem of running enormous deep learning models on constrained edge devices by splitting them into modular fragments~\cite{shi2020communication, lim2020federated}. However, in Cloud-AI where AI systems are deployed on cloud machines, the high communication latency leads to high average response times, making it unsuitable for latency-critical applications like healthcare, gaming and augmented reality~\cite{kraemer2017fog, siriwardhana2021survey, tuli2020healthfog}. Instead, Edge-AI provides low-latency service delivery, thanks to edge devices being in the same Local Area Network (LAN), where the input data from multiple edge nodes are combined to a \blue{single fixed broker node} for processing. Edge-AI based methods aim at scheduling deep neural networks for providing predictable inference~\cite{xiao2018gandiva, gujarati2020serving}. However, due to the centralized collection of data, these solutions typically suffer from high bandwidth overheads and poor service quality~\cite{shi2020communication}. Federated learning depends on data distribution over multiple nodes where the model training and inference are performed in a decentralized fashion. However, this paradigm assumes that neural models with data batches can be accommodated in the system memory. This is seldom the case for common edge devices like Arduinos or Raspberry Pis~\cite{chen2019deep}.

Other recent works offer lower precision models that can fit within the limited memory of such devices by using methods like Model Compression or Model Pruning~\cite{capotondi2020cmix, gunasekaran2020implications, huang2020clio}. However, compressed and low-precision models lose inference accuracy, making them unsuitable for accuracy-sensitive applications like security and intrusion detection~\cite{le2020overview}. Recently, split neural network models have been proposed. They show that  using semantic or layer-wise splitting, a large deep neural network can be fragmented into multiple smaller networks for dividing network parameters onto multiple nodes \cite{matsubara2019distilled, ushakov2018split, kim2017splitnet, gordon2008self}. The former partitions a neural network into parallel disjoint models that produce a part of the result. The latter partitions a neural network into sequential models that generate intermediate results. We illustrate the accuracy and response time trade-offs through sample test cases in Section~\ref{sec:motivation}. Our experiments show that using layer and semantic splitting gives higher inference accuracies than previously proposed model compression techniques (see Section~\ref{sec:experiment}). However, no appropriate scheduling policies exist that can intelligently place such modular neural fragments on a distributed infrastructure to optimize both accuracy and SLA together. The placement of such split models is non-trivial considering the diverse and complex dynamism of task distribution, model usage frequencies and geographical placement of mobile edge devices~\cite{ahmed2017mobile}.
 
% Placement is challenging
\textbf{Research Contributions.} This work proposes a novel neural splitting and placement policy, \textit{SplitPlace}, for enhanced distributed neural network inference at the edge.  SplitPlace leverages a mobile edge computing platform to achieve low latency services. It allows modular neural models to be integrated for best result accuracies that could only be provided by cloud deployments. SplitPlace is the \textit{first} splitting policy that dynamically decides between semantic and layer-wise splits to optimize both inference accuracy and the SLA violation rate. This decision is taken for each incoming task and remains unmodified until the execution of all split fragments of that task are complete. The idea behind the proposed splitting policy is to decide for each incoming task whether to use the semantic or layer-wise splitting strategy based on its SLA demands. Due to their quick adaptability, SplitPlace uses Multi-Armed-Bandits to model the decision strategy for each application type by checking if the SLA deadline is higher or lower than an estimate of the response time for a layer split decision~\cite{bouneffouf2020survey}. Further, SplitPlace optimizes the placement decision of the modular neural network fragments using a split decision aware surrogate model. Compared to a preliminary extended abstract of this work~\cite{tuli2021splitplace}, this paper provides a substantially expanded exposition of the working of MABs in SplitPlace. We also present techniques to dynamically adapt to non-stationary workloads and mobile environments. We present a gradient-based optimization approach for task placement decision conditioned on split decisions.  Experiments on real-world application workloads on a physical edge testbed show that the SplitPlace approach outperforms the baseline approaches by reducing the SLA violation rate and improving the average inference accuracy. 

\textbf{Outline.} The rest of the paper presents a brief background with motivation and related work in Section~\ref{sec:related}. Sections~\ref{sec:model} presents the system model assumptions and formulates the problem. Sections~\ref{sec:splitplace} and~\ref{sec:implementation} give the model details of the proposed SplitPlace approach. We then validate and show the efficacy of the placement policy in Section~\ref{sec:experiment}. Finally, Section~\ref{sec:conclusion} concludes the work and proposes future directions. Additional experimental results are given in the Appendix~\ref{sec:appendix} in the supplementary text.

% \begin{figure}[!t]
%     \centering
%     \subfigure[Semantic split execution]{
%     \includegraphics[width=0.45\linewidth]{images/sem-arch.pdf}
%     \label{fig:sem}
%     }
%     \subfigure[Layer split execution]{
%     \includegraphics[width=0.45\linewidth]{images/layer-arch.pdf}
%     \label{fig:layer}
%     }
%     \caption{Execution pipelines for layer and semantic splitting strategies.}
%     \label{fig:scale}
% \end{figure}

\section{Background and Related Work}
\label{sec:motivation}
\label{sec:related}

As discussed in Section~\ref{sec:introduction}, there is a need for frameworks that can exploit the low latency of edge nodes and also high inference performance of DNNs to provide HALL services. However, complete neural models with the input batch can seldom be accommodated in the random-access-memory (RAM) of edge devices. Thus, ideas like model compression or splitting are required to make inference over large-scale neural networks plausible in such environments. Frameworks that aim at achieving this must maintain a careful balance between accuracy requirements and response times for different user tasks. For such a framework, real-time analysis of the incoming tasks is required for quick decision making of task placement. This requires robust algorithms to seamlessly integrate different paradigms and meet the user's service level agreements.  

\textbf{Semantic and Layer Splitting.} In this work, we leverage the only two available splitting schemes for neural networks: layer and semantic splitting~\cite{gillis, kim2017splitnet}. An overview of these two strategies is shown in Figure~\ref{fig:scale}. Semantic splitting divides the network weights into a hierarchy of multiple groups that use a different set of features (different colored models in Figure~\ref{fig:scale}). Here, the neural network is split based on the data semantics, producing a tree structured model that has no connection among branches of the tree, allowing parallelization of input analysis~\cite{kim2017splitnet}. Due to limited information sharing among the neural network fragments, the semantic splitting scheme gives lower accuracy in general. Semantic splitting requires a separate training procedure where publicly available pre-trained models cannot be used. This is because a pre-trained standard neural network can be split layer wise without affecting output semantics. For semantic splitting we would need to first split the neural network based on data semantics and re-train the model. However, semantic splitting provides parallel task processing and hence lower inference times, more suitable for mission-critical tasks like healthcare and surveillance. Layer-wise splitting divides the network into groups of layers for sequential processing of the task input, shown as different colored models in Figure~\ref{fig:scale}. 

\begin{figure}
    \centering
    \includegraphics[width=\linewidth]{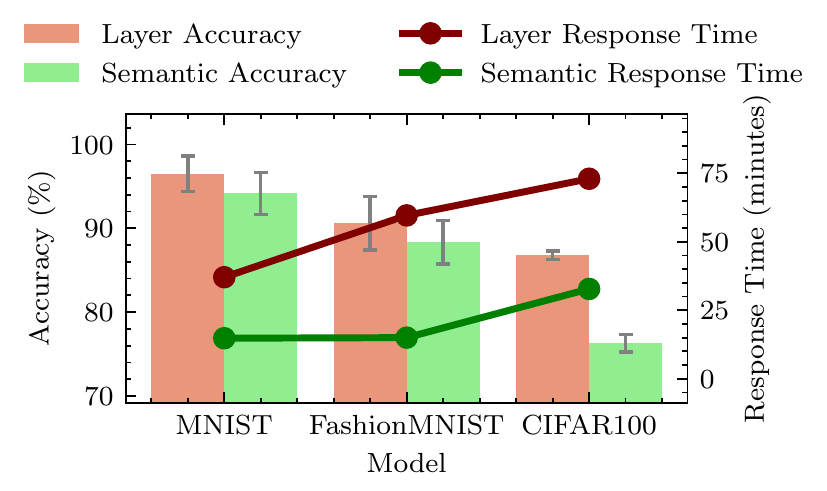}
    \caption{Comparison of Layer and Semantic Splits.}
    \label{fig:layer_sem}
\end{figure}
Layer splitting is easier to deploy as pre-trained models can be just divided into multiple layer groups and distributed to different mobile edge nodes. However, layer splits require a semi-processed input to be forwarded to the subsequent edge node with the final processed output to be sent to the user, thus increasing the overall execution time. Moreover, layer-wise splitting gives higher accuracy compared to semantic splitting. Comparison of accuracies and average response times for the two strategies is shown in Figure~\ref{fig:layer_sem}. The figure shows results for 10 edge \blue{worker nodes} using popular image classification datasets: MNIST, FashionMNIST and CIFAR100~\cite{lecun1998gradient, xiao2017fashion, krizhevsky2009learning} averaged over ResNet50-V2, MobileNetV2 and InceptionV3 neural models~\cite{gunasekaran2020implications}.   As is apparent from the figure, layer splits provide higher accuracy and response time, whereas semantic splits provide lower values for both. SplitPlace leverages this contrast in traits to trade-off between inference accuracy and response time based on SLA requirements of the input tasks. Despite the considerable drop in inference accuracy when using semantic splitting scheme, it is still used in the proposed SplitPlace approach as it is better than model compression or early-exit strategies for quick inference. This is acceptable in many industrial applications~\cite{goli2020migrating, huang2020clio} where latency and service level agreements are more important performance metrics than high-fidelity result delivery. In this work, we consider a system with both SLA violation rates and inference accuracy as optimization objectives. This makes the combination of layer and semantic splitting a promising choice for such use cases.

\begin{table*}[t]
    \centering
    \caption{Comparison of related works with different parameters (\checkmark means that the corresponding feature is present).}
    \resizebox{\textwidth}{!}{
    \begin{tabular}{@{}lcccccccccc@{}}
    \toprule 
    \multirow{2}{*}{Work} & Edge & \multirow{2}{*}{Mobility} & Heterogeneous & Adaptive & Layer & Semantic & Model & \multicolumn{3}{c}{Optimization Parameters}\tabularnewline
    \cline{9-11}
     & Only &  & Environment & QoS & Split & Split & Compression & Accuracy & SLA & Reward\tabularnewline
    \midrule
    \cite{eshratifar2019bottlenet, shao2020bottlenet++} &  &  &  &  &  &  & \checkmark &  & \checkmark & \tabularnewline
    \cite{gao2020discrete} & \checkmark &  & \checkmark &  &  &  & \checkmark & \checkmark &  & \tabularnewline
    \cite{matsubara2019distilled, teerapittayanon2017distributed, goli2020migrating, kang2017neurosurgeon} &  & \checkmark &  &  & \checkmark &  &  &  & \checkmark & \tabularnewline
    \cite{gunasekaran2020implications} & \checkmark &  & \checkmark & \checkmark & \checkmark &  &  & \checkmark & \checkmark & \checkmark\tabularnewline
    \cite{liang2020ai, zhang2021deepslicing} & \checkmark &  &  & \checkmark & \checkmark &  &  &  &  & \tabularnewline
    \cite{kim2017splitnet, ushakov2018split, goli2020migrating, huang2020clio} & \checkmark &  & \checkmark & \checkmark &  & \checkmark &  & \checkmark & \checkmark & \tabularnewline
    \cite{gillis} & \checkmark &  & \checkmark & \checkmark & \checkmark &  & \checkmark & \checkmark & \checkmark & \tabularnewline
    \textbf{SplitPlace} & \checkmark & \checkmark & \checkmark & \checkmark & \checkmark & \checkmark &  & \checkmark & \checkmark & \checkmark\tabularnewline
    \bottomrule 
    \end{tabular}
    }
    \label{tab:related_works}
\end{table*}

\subsection{Related Work}

We now analyze the prior work in more detail. We divide our literature review into three major sections based on the strategy used to allow DNN inference on resource-constrained mobile-edge devices: model compression, layer splitting and semantic splitting. Moreover, we compare prior work based on whether they are suitable for edge-only setups (\textit{i.e.}, without leveraging cloud nodes), consider heterogeneous and mobile nodes and work in settings with adaptive Quality of Service (QoS). See Table~\ref{tab:related_works} for an overview.

\textbf{Model Compression:} Efficient compression of DNN models has been a long studied problem in the literature~\cite{deng2020model}. Several works have been proposed that aim at the structural pruning of neural network parameters without significantly impacting the model's performance. These use approaches like tensor decomposition, network sparsification and data quantization~\cite{deng2020model}. Such pruning and model compression approaches have also been used by the systems research community to allow inference of massive neural models on devices with limited resources~\cite{chandakkar2017strategies}.  Recently, architectures like BottleNet and Bottlenet++ have been proposed~\cite{eshratifar2019bottlenet, shao2020bottlenet++} to enable DNN inference on mobile cloud environments and reduce data transmission times.  BottleNet++ compresses the intermediate layer outputs before sending them to the cloud layer. It uses a model re-training approach to prevent the inference being adversely impacted by the lossy compression of data. Further, BottleNet++ classifies workloads in terms of compute, memory and bandwidth bound categories and applies an appropriate model compression strategy. Other works propose to efficiently prune network channels in convolution neural models using reinforcement learning~\cite{gao2020discrete, wang2020context}. Other efforts aim to prune the weights of the neural models to minimize their memory footprint~\cite{yu2020easiedge}. Such methods aim at improving the accuracy per model size as a metric in contrast to the result delivery time as in BottleNet++. However, model compression does not leverage multiple compute nodes and has poor inference accuracy in general compared to semantic split execution (discussed in Section~\ref{sec:experiment}). Thus, SplitPlace does not use the model compression technique.

\textbf{Layer Splitting:} Many other efforts aim at improving the inference time or accuracy by efficient splitting of the DNN models. Some methods aim to split the networks layer-wise or vertically, \textit{viz}, that the different fragments correspond to separate layer groups and hence impose the constraint of sequential execution. Most work in this category aims at segregating these network splits into distinct devices based on their computational performance~\cite{matsubara2019distilled, teerapittayanon2017distributed, goli2020migrating, zhang2021deepslicing, kang2017neurosurgeon}. In heterogeneous edge-cloud environments, it is fairly straightforward to split the network into two or three fragments each being deployed in a mobile device, edge node or a cloud server. Based on the SLA, such methods provide early-exits if the turnaround time is expected to be more than the SLA deadline. This requires a part of the inference being run at each layer of the network architecture instead of traditionally executing it on the cloud server. Other recent methods aim at exploiting the resource heterogeneity in the same network layer by splitting and placing DNNs based on user demands and \blue{edge worker} capabilities~\cite{gunasekaran2020implications}. Such methods can not only split DNNs, but also choose from different architectural choices to reach the maximum accuracy while agreeing to the latency constraints. Other works aim at accelerating the model run-times by appropriate scheduling of a variety of DNN models on edge-clusters~\cite{liang2020ai}. The state-of-the-art method, \textit{Gillis} uses a hybrid model, wherein it employs either model-compression or layer-splitting based on the application SLA demands~\cite{gillis}. The decision is taken using a reinforcement-learning model which continuously adapts in dynamic scenarios. As the model paritioning is also performed dynamically, the Gillis model cannot work with semantic splitting strategy that requires a new model to be trained for each partitioning scheme. It is a serverless based model serving system that automatically partitions a large model across multiple serverless functions for faster inference and reduced memory footprint per function. The Gillis method employs two model partitioning algorithms that respectively achieve latency optimal serving and cost-optimal serving with service-level agreement compliance. However, this method cannot jointly optimize both latency and SLA. Moreover, it does not consider the mobility of devices or users and hence is ineffective in efficiently managing large DNNs in mobile edge computing environments. 

\textbf{Semantic Splitting:} The early efforts of semantic splitting only split the neural network at the input layer to allow model parallelization and size reduction~\cite{gordon2008self}. Some methods divide the data batch itself across multiple nodes addressing computational contention problems but not memory limitations of fitting neural networks in the RAM~\cite{kaplunovich2020automatic}. Other methods use progressive slicing mechanisms to partition neural models into multiple components to fit in heterogeneous devices~\cite{huang2020clio}. Such methods produce the complete output from each split or fragment of the DNN, adversely impacting the scalability of such methods to high-dimensional output spaces such as image segmentation applications~\cite{kim2017splitnet, ushakov2018split}. Moreover, these methods require significant cross-communication among network splits, significantly increasing the communication overheads. Recently, more intelligent approaches have been developed which hierarchically split neural networks such that each fragment produces a part of the output using an intelligently chosen sub-part of the input~\cite{kim2017splitnet}. Such schemes use the "semantic" information of the data to create the corresponding links between input and output sub-parts being given to each DNN fragment, hence the name \textit{semantic splitting}. Such splitting schemes require minimal to no interaction among network fragments eliminating the communication overheads and increased latency due to stragglers. As semantic splitting can provide results quickly, albeit with reduced accuracy, SplitPlace uses it for tasks with tight deadlines (Section~\ref{sec:splitplace}).

\section{System Model and Problem Formulation}
\label{sec:model}

\begin{figure*}
    \centering %\setlength{\belowcaptionskip}{-10pt}
    \includegraphics[width=0.85\linewidth]{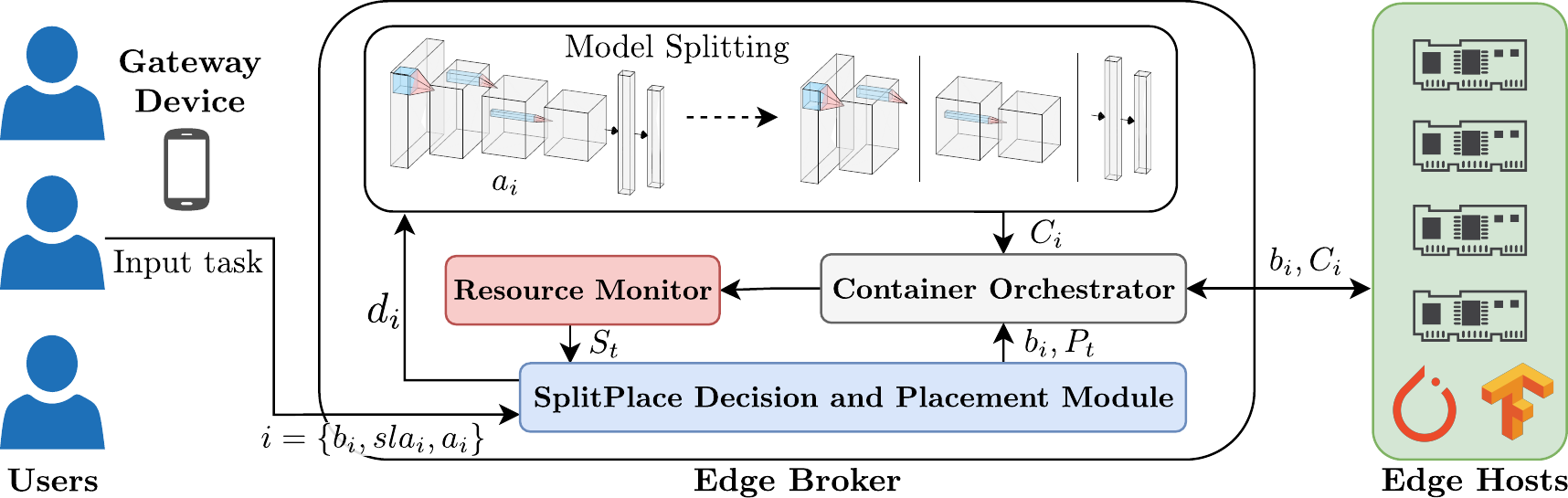}
    \caption{SplitPlace System Model}
    \label{fig:model}
\end{figure*}

In this work, we assume a scenario with a fixed number of multiple heterogeneous edge nodes in a broker-worker fashion, which is a typical case in mobile-edge environments~\cite{tuli2019fogbus, basu2019learn, shao2020bottlenet++, matsubara2019distilled}. \blue{Here, the broker node takes all resource management related decisions, such as neural network splitting and task placement. The processing of such tasks is carried out by the worker nodes.} Examples of broker nodes include personal laptops, small-scale servers and low-end workstations~\cite{tuli2019fogbus}. \blue{Example of common worker nodes in edge environments include Raspberry Pis, Arduino and similar System-on-Chip (SoC) computers~\cite{gill2019transformative}. All tasks are received from an IoT layer that includes sensors and actuators to collect data from the users and send it to the edge broker via the gateway devices. Akin to typical edge configurations~\cite{tuli2021cosco}, the edge broker then decides which splitting strategy to use and schedules these fragments to various edge nodes based on deployment constraints like sequential execution in a layer-decision. The data to be processed comes from the IoT sensors/actuators, which with the decision of which split fragment to use is forwarded by the broker  to each worker node. Some worker nodes are assumed to be mobile, whereas others are considered to be fixed in terms of their geographical location. In our formulation, we consider mobility only in terms of the variations in terms of the network channels and do not consider the worker nodes or users crossing different networks. We assume that the CPU, RAM, Bandwidth and Disk capacities of all nodes are known in advance, and similarly the broker can sample the resource consumption for each task in the environment at any time (see \textit{Resource Monitor} in Fig.~\ref{fig:model}). As we describe later, the broker periodically measure utilizations of CPU, RAM, Bandwidth and Disk for each task in the system. The broker is trusted with this information such that it can make informed resource management decisions to optimize QoS.} Moreover, we consider that tasks include a batch of inputs that need to be processed by a DNN model. Further, for each task, a service level deadline is defined at the time the task is sent to the edge environment.  We give an overview of the SplitPlace system model in Figure~\ref{fig:model}. We decompose the problem into deciding an optimal splitting strategy and a fragment placement for each application (motivation in Appendix~\ref{app:splitplace} and more details in Section~\ref{alg:splitplace}).

\textbf{Workload Model.} We consider a bounded discrete time control problem where we divide the timeline into equal duration intervals, with the $t$-th interval denoted as $I_t$. \blue{Here, $t \in \{0, \ldots, \Gamma\}$, where $\Gamma+1$} is the number of intervals in an execution. We assume a fixed number of \blue{worker} machines in the edge layer and denote them as $\mathcal{H}$. We also consider that new tasks created at the interval $I_t$ are denoted as $\mathcal{N}_t$, with all active tasks being denoted as $\mathcal{T}_t$ (and $\mathcal{N}_t \subseteq \mathcal{T}_t$). Each task $i \in \mathcal{T}_t$ consists of a batch input $b_i$, SLA deadline $sla_i$ and a DNN application $a_i$. The set of all possible DNN applications is denoted by $\mathcal{A}$. For each new task $i \in \mathcal{N}_t$, the edge broker takes a decision $d^i$, such that $d^i \in \{L, S\}$, with $L$ denoting layer-wise splitting and $S$ denoting semantic split strategy. The collection of all split decisions for active tasks in interval $I_t$ is denoted as $\mathcal{D}_t = \{d^i\}_{i \in \mathcal{N}_t}$. Based on the decision $d^i$ for task $i$, this task is realized as an execution workflow in the form of containers $C^i$. Similar to a VM, a container is a package of virtualized software that contains all of the necessary elements to run in any environment. The set of all containers active in the interval $I_t$ is denoted as $C_t = \cup_{i \in \mathcal{T}_t} C^i$. The set of all utilization metrics of CPU, RAM, Network Bandwidth and Disk for all containers and \blue{workers} at the start of the interval $I_t$ defines the state of the system, denoted as $S_t$. A summary of the symbols is given in Table~\ref{tab:symbols}.

\begin{table*}[]
    \centering
    \caption{Table of Main Notation}
    \resizebox{0.7\linewidth}{!}{
    \begin{tabular}{@{}cc@{}}
        \toprule
        Notation & Description \\
        \midrule
        $I_t$ & $t$-th scheduling interval \\ 
        \blue{$\mathcal{T}_t$} & Active tasks in $I_t$ \\ 
        \blue{$\mathcal{N}_t$} & New tasks received at the start of $I_t$ \\ 
        \blue{$\mathcal{H}$} & Set of \blue{workers} in the edge layer \\ 
        $i = \{b_i, sla_i, a_i\}$ & Task as a collection input batch, SLA deadline and application type \\
        $d^i \in \{L, S\}$ & Splitting decision for input task $i$ \\ 
        $C^i$ & Container realization of task $i$ based on decision $d^i$\\ 
        $C_t = \cup_{i \in \mathcal{T}_t} C^i$ & Set of all active containers in the interval $I_t$\\ 
        $S_t$ & State of the system at the start of $I_t$\\ 
        $P_t : C_t \times \mathcal{H}$  & Placement decision of $C_t$ to $\mathcal{H}$ as an adjacency matrix\\ 
        $O_t$  & Objective score for interval $I_t$\\ \bottomrule
    \end{tabular}
    }
    \label{tab:symbols}
\end{table*}

\subsection{Split Nets Placement}

\begin{figure}
    \centering %\setlength{\belowcaptionskip}{-10pt}
    \includegraphics[width=0.85\linewidth]{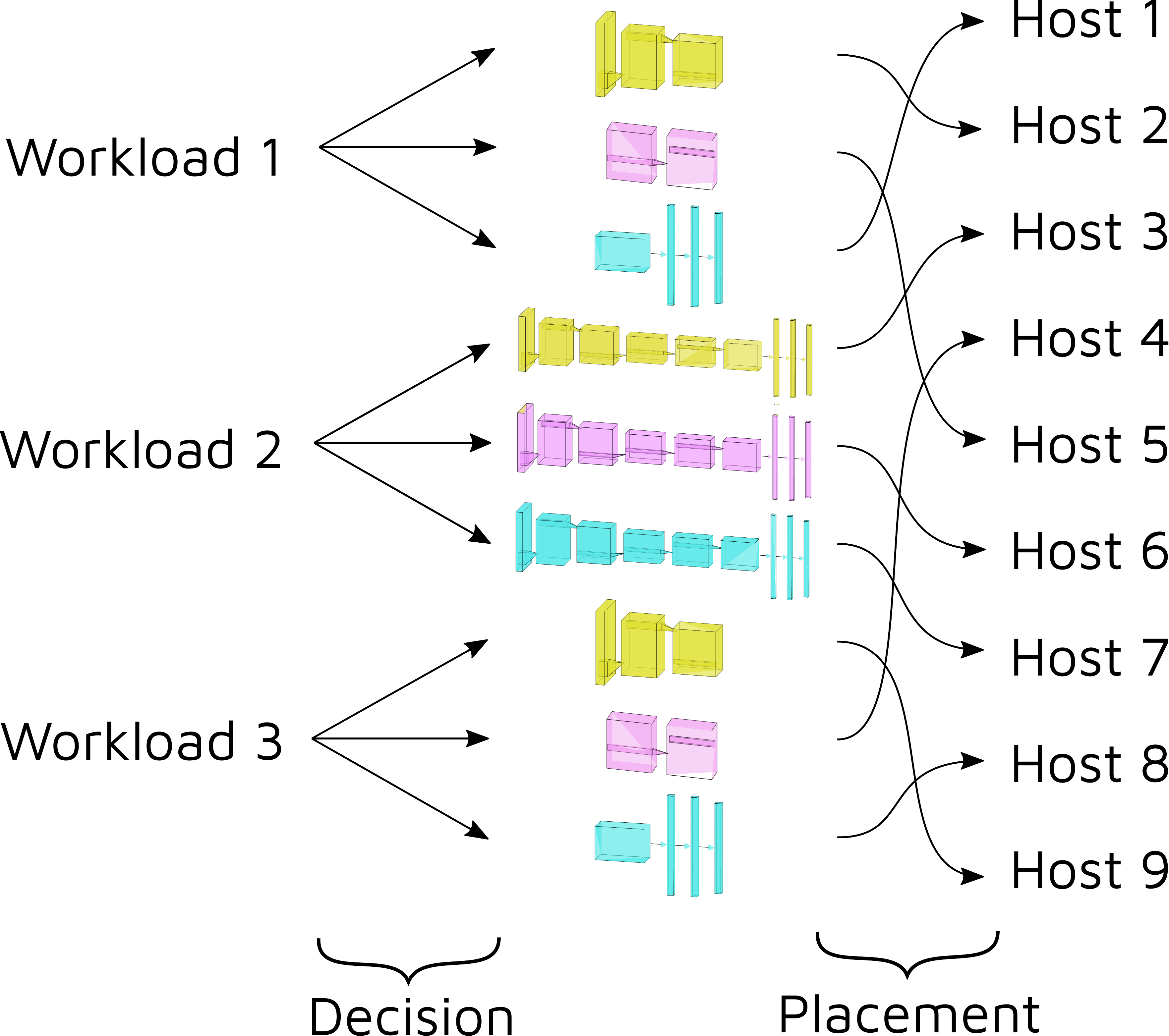}
    \caption{SplitPlace decision and placement problems.}
    \label{fig:problems}
\end{figure}

We partition the problem into two sub-problems of deciding the optimal splitting strategy for input tasks and that of placement of active containers in edge \blue{workers} (see Figure~\ref{fig:problems}). Considering the previously described system model, at the start of each interval $I_t$, the SplitPlace model takes the split decision $d^i$ for all $i \in \mathcal{N}_t$. Moreover, it also takes a placement decision for all active containers $C_t$, denoted as an adjacency matrix $P_t : C_t \times \mathcal{H}$. This is realized as a container allocation for new tasks and migration for active tasks in the system. 

The main idea behind the layer-wise split design is first to divide neural networks into multiple independent splits, classify these splits in preliminary, intermediate and final neural network layers and distribute them across different nodes based on the node capabilities and network hierarchy. This exploits the fact that communication across edge nodes in the LAN with few hop distances is very fast and has low latency and jitter \cite{park2016joint}. Moreover, techniques like knowledge distillation can be further utilized to enhance the accuracy of the results obtained by passing the input through these different classifiers. However, knowledge distillation needs to be applied at the training stage, before generating the neural network splits. As there are many inputs in our assumed large-scale deployment, the execution can be performed in a pipelined fashion to further improve throughput over and above the low response time of the nodes at the edge of the network. For the semantic split, we divide the network weights into a set or a hierarchy of multiple groups that use disjoint sets of features. This is done by making assignment decisions of network parameters to edge devices at deployment time. This produces a tree-structured network that involves no connection between branched sub-trees of semantically disparate class groups. Each sub-group is then allocated to an edge node. The input is either broadcasted from the broker or forwarded in a ring-topology to all nodes with the network split corresponding to the input task. We use standard layer~\cite{gillis} and semantic splitting~\cite{kim2017splitnet} methods as discussed in Section~\ref{sec:related}.

We now outline the working of the proposed distributed deep learning architecture for edge computing environments. Figure~\ref{fig:model} shows a schematic view of its working. As shown, there is a shared repository of neural network parameters which is distributed by the broker to multiple edge nodes. The layer and semantic splits are realized as Docker container images that are shared by the broker to the worker nodes at the start of each execution trace. The placement of tasks is realized as spinning up a Docker container using the corresponding image on the worker. \blue{As the process of sharing container images is a one-time event, transferring all semantic and layer split fragments does not impose a high overhead on network bandwidth at runtime. This sharing of containers for each splitting strategy and dataset type are transferred to the worker nodes is performed at the start of the run. At run-time, only the decision of which split fragment to be used is communicated to the worker nodes, which executes a container from the corresponding image.} The placement of task on each worker is based on the resource availability, computation required to be performed in each section and the capabilities of the nodes (obtained by the \textit{Resource Monitor}). For intensive computations with large storage requirements (Gated Recurrent Units or LSTMs) or splits with high dimension size of input/output (typically the final layers), the splits are sent to high-resource edge \blue{workers}. The management of allocation and migration of neural network splits is done by the \textit{Container Orchestrator}. Other attention based sub-layer extensions can be deployed in either edge or cloud node based on application requirements, node constraints and user demands. Based on the described model assumptions, we now formulate the problem of taking splitting and placement decisions to optimize the QoS parameters. Implementation specific details on how the results of layer-splits are forwarded and outputs of semantic splits combined across edge nodes are given in Section~\ref{sec:implementation}.

\subsection{Problem Formulation}
\label{sec:formulation}

The aim of the model is to optimize an objective score $O_t$ (to be maximized), which quantifies the QoS parameters of the interval $I_t$, such as accuracy, SLA violation rate, energy consumption and average response time. \blue{This is typically in the form of a convex combination of energy consumption, response time, SLO violation rates, etc.~\cite{tuli2021cosco, da2018resource}.} The constraints in this formulation include the following. Firstly, the container decomposition for a new task $i \in \mathcal{N}_t$ should be based on $d^i$. Secondly, containers corresponding to the layer-split decisions $\{C^i | d^i = L\}$ should be scheduled as per the linear chain of precedence constraints. This means that a container later in the neural inference pipeline should be scheduled only after the complete execution of the previous containers in the pipeline. This is because the output of an initial layer in an inference pipeline of a neural network is required before we can schedule a latter layer in the pipeline. Thirdly, the placement matrix $P_t : C_t \times \mathcal{H}$ should adhere to the allocation constraints, \textit{i.e.}, it should not allocate/migrate a container to a \blue{worker} where the \blue{worker} does not have sufficient resources available to accommodate the container. Thus, the problem can be formulated as
\begin{equation}
\label{eq:problem}
\begin{aligned}
& \underset{P_t, \mathcal{D}_t}{\text{maximize}}
& & \sum_{t}^{T} O_t \\
& \text{subject to}
& & \forall\ t, \forall\ i \in \mathcal{N}_t, C^i \text{ containers created } \\ & & & \text{based on splitting decision } d^i, \\
&&& \forall\ t, P_t \text{ is feasible}, \\
&&& \forall\ d^i = L, C^i \text{ follow precedence chain}.
\end{aligned}
\end{equation}

\begin{figure*}
    \centering
    \includegraphics[width=0.9\linewidth]{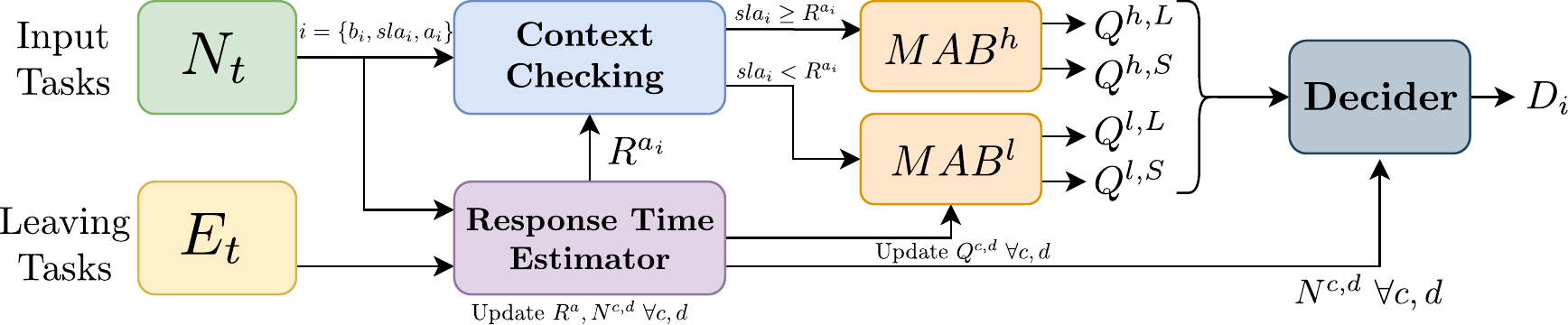}
    \caption{MAB decision workflow.}
    \label{fig:decider}
\end{figure*}

\section{SplitPlace Policy}
\label{sec:splitplace}

We now describe the SplitPlace decision and placement policy. For the first sub-problem of deciding optimal splitting strategy, we employ a Multi-Armed Bandit model to dynamically enforce the decision using external reward signals\footnote{Compared to other methods like A/B testing and Hill Climbing search~\cite{kohavi2017online}, Multi-Armed Bandits allow quick convergence in scenarios when different cases need to be modelled separately, which is the case in our setup. Thus, we use Mult-Armed Bandits for deciding the optimal splitting strategy for an input task.}. Our solution for the second sub-problem of split placement uses a reinforcement-learning based approach that specifically utilizes a surrogate model to optimize the placement decision (agnostic to the specific implementation)\footnote{In contrast to Monte Carlo or Evolutionary methods, Reinforcement learning allows placement to be goal-directed, \textit{i.e.}, aims at optimizing QoS using it as a signal, and allows the model to adapt to changing environments~\cite{wiering2012reinforcement}. Hence, we use a RL model, specifically using a surrogate model due for its scalability, to decide the optimal task placement of network splits.}. \blue{This two-stage approach is suboptimal since the response time of the splitting decision depends on the placement decision. In case of large variation in terms of the computational resources, it is worth exploring joint optimization of both decisions. However, in our large-scale edge settings, this segregation helps us to make the problem tractable as we describe next.} 

The motivation behind this segregation is two-fold. First, having a single reinforcement-learning (RL) model that takes both splitting and placement decisions makes the state-space explode exponentially, causing memory bottlenecks in resource-constrained edge devices~\cite{chen2018optimized}. Having a simple RL model does not allow it to scale well with several devices in modern IoT settings (see Section~\ref{sec:experiment} with 50 edge devices the and Gillis RL baseline). One of the solutions that we explore in this work is to simplify this complex problem by decomposing it into split decision making and task placement. Second, the response time of an application depends primarily on the splitting choice, layer or semantic, making it a crucial factor for SLA deadline based decision making. To minimize the SLA violation rates we only use the response time based context for our Multi-Armed bandit model. Other parameters like CPU or RAM utilization have high variability in a volatile setting and are not ideal choices for deciding which splitting strategy to opt. Instead, the inference accuracy is another key factor in taking this decision. Thus, SLA violation and inference accuracy are apt objectives for the first sub-problem. Further, the energy consumption and average response time largely depend on the task placement, making them an ideal objective for optimization in the task placement sub-problem.

\subsection{Multi-Armed Bandit Decision Module}

Multi-Armed Bandit, in short MAB, is a policy formulation where a state-less agent is expected to take one of many decisions with each decision leading to a different reward. The objective of such an agent is to maximize the expected long-term reward~\cite{bouneffouf2020survey}. However, in our case, the most important factor to consider when making a decision of whether to use layer or semantic splits for a task is its SLA deadline. 

\subsubsection{Estimating Response Time of Layer-Splits}

The idea behind the proposed SplitPlace approach is to maintain MABs for two different contexts: 1) when SLA is greater than the estimate of the response time for a layer decision, 2) when SLA is less than this estimate. The motivation behind these two contexts is that in case of the SLA deadline being lower than the execution time of layer split, a "layer" decision would be more likely to violate the SLA as result delivery would be after the deadline. However, the exact time it takes to completely execute all containers corresponding to the layer split decision is apriori unknown. Thus, for every application type, we maintain estimates of the response time, \textit{i.e}, the total time it takes to execute all containers corresponding to this decision. 

Let us denote the tasks leaving the system at the end of $I_t$ as $E_t$. Now, for each task $i \in E_t$, we denote response time and inference performance using $r_i$ and $p_i$. We denote the layer response time estimate for application $a \in \mathcal{A}$ as $R^a$. To quickly adapt to non-stationary scenarios, for instance due to the mobility of edge nodes in the system, we update our estimates using new data-points as exponential moving averages using the multiplier $\phi \in [0, 1]$ for the most recent response time observation. Moving averages presents a low computational cost and consequently low latency compared to more sophisticated smoothing functions.
\begin{equation}
    \label{eq:r_update}
   R^a \gets \phi \cdot r_i + (1 - \phi) \cdot R^a,\ \forall i \in E_t \wedge d^i = L,\ \forall a \in A.
\end{equation}
Compared to simple moving average, the above equation gives higher weights to the latest response times, allowing the model to quickly respond to recent changes in environment and workload characteristics.

\subsubsection{Context based MAB Model}

Now, for any input task $i \in \mathcal{N}_t$, we divide it into two cases: $sla_i \geq R^{a_i}$ and $sla_i < R^{a_i}$. Considering that the response time of a semantic-split decision would likely be lower than the layer-split decision, in the first case both decisions would most likely not lead to an SLA violation (\textit{high SLA} setting). However, in the second case, a layer-split decision would likely lead to an SLA violation but not the semantic-split decision (\textit{low SLA} setting). To tackle the problem for these different contexts, we maintain two independent MAB models denoted as $MAB^h$ and $MAB^l$. The former represents a MAB model for the high-SLA setting and the latter for the low-SLA setting.

For each context and decision $d \in \{L, S\}$, we define reward metrics as
\begin{gather}
    \label{eq:ohd}
    O^{h, d} = \frac{\sum_{i \in E_t} \left( \mathbbm{1}(r_i \leq sla_i) + p_i \right) \cdot \mathbbm{1}(sla_i \geq R^{a_i} \wedge d^i = d)}{2 \cdot \sum_{i \in E_t} \mathbbm{1}(sla_i \geq R^{a_i} \wedge d^i = d)},\\
    \label{eq:old}
    O^{l, d} = \frac{\sum_{i \in E_t} \left( \mathbbm{1}(r_i \leq sla_i) + p_i \right) \cdot \mathbbm{1}(sla_i < R^{a_i} \wedge d^i = d)}{2 \cdot \sum_{i \in E_t} \mathbbm{1}(sla_i < R^{a_i} \wedge d^i = d)}.
\end{gather}
The first term of the numerator, \textit{i.e.}, $\mathbbm{1}(r_i \leq sla_i)$ quantifies SLA violation reward (one if not violated and zero otherwise). The second term, \textit{i.e.}, $p_i$ corresponds to the inference accuracy of the task. These two objectives have been motivated at the start of Section~\ref{sec:splitplace}. Thus, each MAB model gets the reward function for its decisions allowing independent training of the two. The weights of the two metrics, \textit{i.e.}, accuracy and SLA violation can be set by the user to modify the relative importance between the metrics as per application requirements. In our experiments, the weight parameters of both metrics are set to be equal based on grid-search, maximizing the average reward. 

Now, for each decision context $c \in \{h, l\}$ and $d \in \{L, S\}$, we maintain a decision count $N^{c,d}$ and a reward estimate $Q^{c,d}$ which is updated using the reward functions $O^{h, d}$ or $O^{l, d}$ as follows
\begin{equation}
    \label{eq:q_update}
    Q^{c,d} \gets Q^{c,d} + \gamma (O^{c, d} - Q^{c,d}),\ \forall d \in \{L, S\},\ \forall c \in \{h, l\}.
\end{equation}
where $\gamma$ is the decay parameter. Thus, each reward-estimate is updated by the corresponding reward metric.

\begin{algorithm*}[!t]
    \begin{algorithmic}[1]
    \Require
    \Statex Pre-trained MAB models $MAB^h, MAB^l$
    \Statex Discounting factor $\gamma \in (0, 1)$
    \Procedure{SplitPlace}{scheduling interval $I_t$}
        \State Get new tasks $\mathcal{N}_t$
        \State Get leaving tasks $E_t$
        \State Calculate $O^{c,d}\ \forall c \in \{h, l\}, d \in \{L, S\}$ using \eqref{eq:ohd} and \eqref{eq:old}
        \State Update gain estimates $Q^{c, d}\ \forall c, d$ using \eqref{eq:q_update} \label{line:q} \Comment{Q update}
        \State Update decision counts $N^{c,d}\ \forall c, d$ \label{line:n} \Comment{Count update}
        \For{$i \in \mathcal{N}_t$}
                \State $i = \{b_i, sla_i, a_i\}$
                \State $d^i \gets
                \begin{cases}
                \argmax_{d \in \{L, S\}} Q^{h,d} + c \sqrt{\frac{\log t}{N^{h,d}}}, & sla_i \geq R^{a_i}\\
                \argmax_{d \in \{L, S\}} Q^{l,d} + c \sqrt{\frac{\log t}{N^{l,d}}}, & sla_i < R^{a_i}
                \end{cases}$ \label{line:ucb} \Comment{UCB based decision}
        \EndFor
        \State $\mathcal{D}_t \gets \{d^i\}_{i \in \mathcal{N}_t}$
        \State Get $S_t$ from resource-monitor \Comment{Current State}
        \State $P_t \gets \textsc{DASO}([S_t, P_{t-1}, \mathcal{D}_t])$ \label{line:placement} \Comment{Placement Decision} 
        \State $O^{MAB} \gets \tfrac{1}{4} \sum_{c \in \{h, l\}} \sum_{d \in \{L, D\}} O^{c, d}$
        \State Fine-tune $\textsc{DASO}$ using $O^P$ calculated using~\eqref{eq:reward_placement} \label{line:tune}\\
        \Return $P_t$
    \EndProcedure
    \end{algorithmic}
\caption{SplitPlace Decision and Placement Module}
\label{alg:splitplace}
\end{algorithm*}

For both these MAB models, we use a parameter-free feedback-based $\epsilon$-greedy learning approach that is known to be versatile in adapting to diverse workload characteristics~\cite{maroti2019rbed}. Unlike other strategies, this is known to scale asymptotically as the long-term Q estimates become exact under mild conditions~\cite[\S~2.2]{sutton2018reinforcement}. To train the model, we take the contextual decision
\begin{equation}
    \label{eq:decision_training}
    d^i = 
    \begin{cases}
    \begin{cases}
    \text{random decision}, & \text{with prob. }\epsilon\\
    \argmax_{d \in \{L, S\}} Q^{h,d}, & \text{otherwise}
    \end{cases}, & sla_i \geq R^{a_i}\\
    \begin{cases}
    \text{random decision}, & \text{with prob. }\epsilon\\
    \argmax_{d \in \{L, S\}} Q^{l,d}, & \text{otherwise}
    \end{cases}, & sla_i < R^{a_i}
    \end{cases}.
\end{equation}
Here, the probability $\epsilon$ decays using the reward feedback, starting from 1. We maintain a reward threshold $\rho$ that is initialized as a small positive constant $k < 1$, and use average reward $O^{MAB} = \tfrac{1}{4} \sum_{c \in \{h, l\}} \sum_{d \in \{L, D\}} O^{c, d}$ to update $\epsilon$ and $\rho$ using the rules
\begin{gather}
\label{eq:decay_increment}
    \epsilon \gets
    \begin{cases}
    decay(\epsilon), & O^{MAB} > \rho\\
    \epsilon, & \text{otherwise}
    \end{cases},\\
    \rho \gets
    \begin{cases}
    increment(\rho), & O^{MAB} > \rho\\
    \rho, & \text{otherwise}
    \end{cases}.
\end{gather}
Here $decay(\epsilon) = (1 - k) \cdot \epsilon$ and $increment(\rho) = (1 + k) \cdot \rho$. Note that $O^{MAB} > \rho$ refers to the current value of $\rho$ prior to the update. The $k$ value controls the rate of convergence of the model. The $\epsilon$ value controls the exploration of the model at training time allowing the model to visit more states and obtain precise estimates of layer-split response times.

However, at test time we already have precise estimates of the response times; thus exploration is only required to adapt in volatile scenarios. For this, $\epsilon$-greedy is not a suitable approach as decreasing $\epsilon$ with time would prevent exploration as time progresses. Instead, we use an Upper-Confidence-Bound (UCB) exploration strategy that is more suitable as it takes decision counts also into account~\cite{zhang2019multi, gupta2020correlated}.  Thus, at test time, we take a deterministic decision using the rule
\begin{equation}
\label{eq:decision_testing}
    d^i = 
    \begin{cases}
    \argmax_{d \in \{L, S\}} Q^{h,d} + c \sqrt{\frac{\log t}{N^{h,d}}}, & sla_i \geq R^{a_i}\\
    \argmax_{d \in \{L, S\}} Q^{l,d} + c \sqrt{\frac{\log t}{N^{l,d}}}, & sla_i < R^{a_i}
    \end{cases},
\end{equation}
where $t$ is the scheduling interval count and $c$ is the exploration factor. An overview of the complete split-decision making workflow is shown in Figure~\ref{fig:decider}. We now discuss the RL based placement module. It is worth noting that both components are independent of each other and can be improved separately in future; however, our experiments show that the MAB decision module accounts for most of the performance gains (see Section~\ref{sec:experiment}).

\subsection{Reinforcement Learning based Placement Module}
\label{sec:placement_module}

Once we have the splitting decision for each input task $i \in \mathcal{N}_t$, we now can create containers using pre-trained layer and semantic split neural networks corresponding to the application $a_i$. This can be done offline on a resource rich system, where the split models can be trained using existing datasets. Once we have trained models, we can generate  container images corresponding to each split fragment and distribute to all worker nodes~\cite{ahmed2018docker}. Then we need to place containers, translating to the worker node initializing a container for the corresponding image, of all active tasks $C_t$ to workers $\mathcal{H}$. To do this, we use a learning model which predicts the placement matrix $P_t$ using the state of the system $S_t$, decisions $\mathcal{D}_t = d^i \forall i \in \mathcal{T}_t$ and a reward signal $O^{P}$. To define $O^P$, we define the following metrics~\cite{tuli2021cosco}:
\begin{enumerate}
    \item \textit{Average Energy Consumption} (AEC) is defined for any interval $I_t$ as the mean energy consumption of all edge \blue{workers} in the system.
    \item \textit{Average Response Time} (ART) is defined for any interval $I_t$ as mean response time (in scheduling intervals) of all leaving tasks $E_t$.
\end{enumerate}

The choice of these two objectives for the placement sub-problem has been motivated at the start of Section~\ref{sec:splitplace}. Using these metrics, for any interval $I_t$, $O^P$ is defined as
\begin{equation}
\label{eq:reward_placement}
    O^P = O^{MAB} - \alpha \cdot AEC_t - \beta \cdot ART_t.
\end{equation}
Here, $\alpha$ and $\beta$ (such that $\alpha+\beta=1$) are hyper-parameters that can be set by users as per the application requirements. Higher $\alpha$ aims to optimize energy consumption at the cost of higher response times, whereas low $\alpha$ aims to reduce average response time. Thus, a RL model $f$, parameterized by $\theta$ takes a decision $P_i$, where the model uses the reward estimate as the output of the function $f([S_t, P_t, \mathcal{D}_t]; \theta)$, where the parameters $\theta$ are updated based on the reward signal $O^P$. We call this learning approach ``decision-aware'' as part of the input is the split-decision taken by the MAB model. 

Clearly, the proposed formulation is agnostic to the underlying implementation of the learning approach. Thus, any policy like Q-learning or Actor-Critic Learning could be used in the SplitPlace model~\cite{sutton2018reinforcement}. However, recently developed techniques like \textit{GOBI}~\cite{tuli2021cosco} use gradient-based optimization of the reward to quickly converge to a local-maximum of the objective function. GOBI uses a neural-network based surrogate model to estimate the reward from a given input state, which is then used to update the state by calculating the gradients of the reward estimates with respect to the input. Moreover, advances like momentum, annealing and restarts allow such models to quickly reach a global optima~\cite{tuli2021cosco}. 

\textbf{DASO placement module.} In the proposed framework, we use decision-aware surrogate based optimization method (termed as DASO) to place containers in a distributed mobile edge environment. This is motivated from prior neural network based surrogate optimization methods~\cite{tuli2021cosco}. Here, we consider a Fully-Connected-Network (FCN) model $f(x; \theta)$ that takes an $x$ as a tuple of input state $S_t$, split-decision $\mathcal{D}_t$ and placement decision $P_t$, and outputs an estimate of the QoS objective score $O_t$. This is because FCNs are agnostic to the structure of the input and hence a suitable choice for modeling dependencies between QoS metrics and model inputs like resource utilization and placement decision~\cite{basu2019learn, tuli2021cosco}. Exploration of other styles of neural models, such as graph neural networks that can take the network topology graph as an input are part of future work. Now, using existing execution trace dataset, $\Lambda = \{ [S_t, P_t, \mathcal{D}_t], O_t \}_b$, the FCN model is trained to optimize its network parameters $\theta$ such that the Mean-Square-Error (MSE) loss
\begin{equation}
    \mathcal{L}(f(x;\theta), y) = \tfrac{1}{b} \textstyle \sum_{t=0}^b (y - f(x;\theta))^2, \text{ where } (x,y) \in \Lambda.
\end{equation}
is minimized as in~\cite{tuli2021cosco}. To do this, we use AdamW optimizer~\cite{loshchilov2018decoupled} and update $\theta$ up till convergence. This allows the surrogate model $f$ to predict an QoS objective score for a given system state $S_t$, split-decisions $\mathcal{D}_t$ and task placement $P_t$. Once the surrogate model is trained, starting from the placement decision from the previous interval $P_t = P_{t-1}$, we leverage it to optimize the placement decision using the following rule
\begin{equation}
    P_t \gets P_t - \eta \cdot \nabla_{P_t}f([S_t, P_t, \mathcal{D}_t]; \theta),
\end{equation}
for a given state and decision pair $S_t, \mathcal{D}_t$. Here, $\eta$ is the learning rate of the model. The above equation is iterated till convergence, \textit{i.e.}, the $L2$ norm between the placement matrices of two consecutive iterations is lower than a threshold value. Thus, at the start of each interval $I_t$, using the output of the MAB decision module, the DASO model gives us a placement decision $P_t$.

\begin{table*}[!t]
    \centering
    \caption{Edge Worker characteristics of Azure Edge Environment.}
    % \resizebox{\linewidth}{!}{
    \begin{tabular}{@{}lccccccccc@{}}
    \toprule 
    \multirow{2}{*}{\textbf{Name}} & \multirow{2}{*}{\textbf{Qty}} & \textbf{Core} & \multirow{2}{*}{\textbf{MIPS}} & \multirow{2}{*}{\textbf{RAM}} & \textbf{RAM} & \textbf{Ping} & \textbf{Network} & \textbf{Disk} & \textbf{Cost}\tabularnewline
     &  & \textbf{count} &  &  & \textbf{Bandwidth} & \textbf{time} & \textbf{Bandwidth} & \textbf{Bandwidth} & \textbf{Model}\tabularnewline
    \midrule
    \multicolumn{10}{c}{\textbf{Worker Nodes}}\tabularnewline
    \midrule
    B2ms & 20 & 2 & 4029 & 4295 MB & 372 MB/s & 2 ms & 1000 MB/s & 13.4 MB/s & 0.0944 \$/hr\tabularnewline
     
    E2asv4 & 10 & 2 & 4019 & 4172 MB & 412 MB/s & 2 ms & 1000 MB/s & 10.3 MB/s & 0.148 \$/hr\tabularnewline
     
    B4ms & 10 & 4 & 8102 & 7962 MB & 360 MB/s & 3 ms & 2500 MB/s & 10.6 MB/s & 0.189 \$/hr\tabularnewline
     
    E4asv4 & 10 & 4 & 7962 & 7962 MB & 476 MB/s & 3 ms & 2500 MB/s & 11.64 MB/s & 0.296 \$/hr\tabularnewline
    \midrule 
    \multicolumn{10}{c}{\textbf{Broker Node}}\tabularnewline
    \midrule
    L8sv2 & 1 & 8 & 16182 & 17012 MB & 945 MB/s & 1 ms & 4000 MB/s & 17.6 MB/s & 0.724 \$/hr\tabularnewline
    \bottomrule 
    \end{tabular}
    % }
    \label{tab:hosts}
\end{table*}

\subsection{SplitPlace Algorithm}

An overview of the SplitPlace approach is given in Algorithm~\ref{alg:splitplace}. Using pre-trained MAB models, \textit{i.e.}, Q-estimates $Q^{c,d}$ and decision counts $N^{c,d}$, the model decides the optimal splitting decision using the UCB metric (line~\ref{line:ucb}). To adapt the model in non-stationary scenarios, we dynamically update the Q-estimates and decision counts (lines~\ref{line:q} and \ref{line:n}). Using the current state and the split-decisions of all active tasks, we use the DASO approach to take a placement decision for the active containers (line~\ref{line:placement}). Again, we fine-tune the DASO's surrogate model using the reward metric to adapt to changes in the environment, for instance the changes in the latency of mobile edge nodes and their consequent effect on the reward metrics (line~\ref{line:tune}). However, the placement decision must conform to the allocation constraints as described in Section~\ref{sec:formulation}. To relax the constraint of having only feasible placement decisions, in SplitPlace we allocate or migrate only those containers for which it is possible. Those containers that could not be allocated in a scheduling interval are placed to nodes corresponding to the highest output of the neural network $f$. If no \blue{worker} placement is feasible the task is added to a wait queue, which are considered again for allocation in the next interval.

\section{Implementation}
\label{sec:implementation}

To implement and evaluate the SplitPlace policy, we need a framework that we can use to deploy containerized neural network split fragments on an edge computing environment. One such framework is COSCO~\cite{tuli2021cosco}. It enables the development and deployment of integrated edge-cloud environments with structured communication and platform independent execution of applications. It connects various IoT sensors, which can be healthcare sensors with gateway devices, to send data and tasks to edge computing nodes, including edge or cloud \blue{workers}. The resource management and task initiation is undertaken on edge nodes in the broker layer. The framework uses HTTP RESTful APIs for communication and seamlessly integrates a \texttt{Flask} based web-environment to deploy and manage containers in a distributed setup~\cite{grinberg2018flask}.

We use only the edge-layer deployment in the framework and use the Docker container engine to containerize and execute the split-neural networks in various edge \blue{workers}~\cite{ahmed2018docker}. We uses the Checkpoint/Restore In Userspace (CRIU)~\cite{venkatesh2019fast} tool for container migration. Further, the DASO approach is implemented using the Autograd package in the PyTorch module~\cite{paszke2017automatic}.

% \subsection{Extending COSCO to handle workflow constraints}

To implement SplitPlace in the COSCO framework, we extend the \texttt{Framework} class to allow constraints for sequential execution of layer-splits. The function \texttt{getPlacementPossible()} was modified to also check for containers of layer-split partitioning scheme to be scheduled sequentially. Moreover, we implemented data transferring pipeline for broadcasting inputs in semantic-split decision and forwarding the outputs in layer-split decision. Finally, the inference outputs were synchronized and brought to the broker to calculate the performance accuracy and measure the workflow response time. For synchronization of outputs and execution of network splits, we use the HTTP Notification API.

%%%%%%%%%%%%%%%%%%%%%
\begin{figure*}[!t]
    \centering
    \subfigure[Average Response Time Estimate]{
    \includegraphics[width=.31\textwidth]{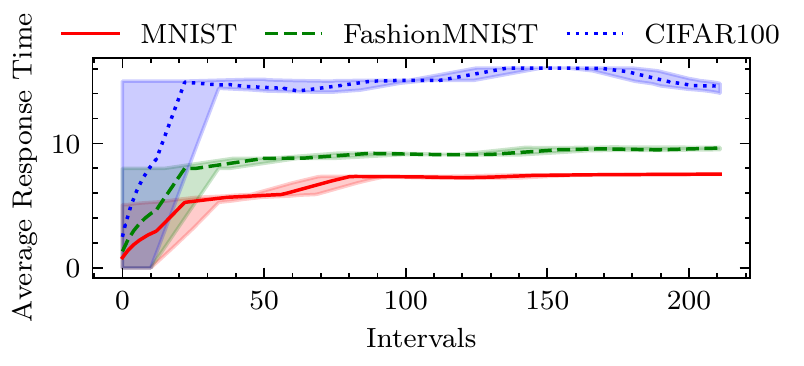}
    \label{fig:mab_response}
    }
    \subfigure[Decision Counts (SLA $>$ estimate)]{
    \includegraphics[width=.31\textwidth]{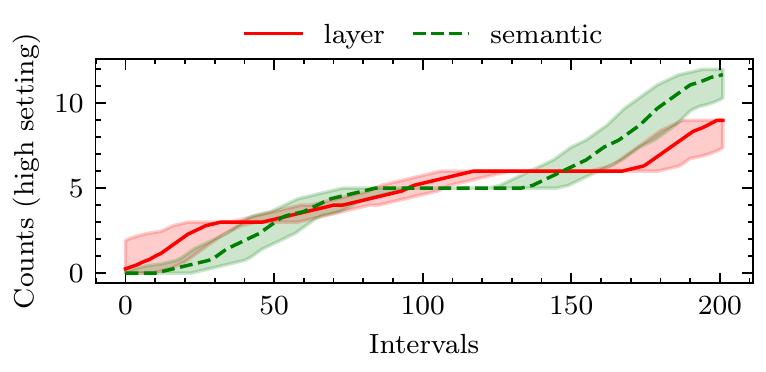}
    \label{fig:mab_counts_high}
    }
    \subfigure[Decision Counts (SLA $<$ estimate)]{
    \includegraphics[width=.31\textwidth]{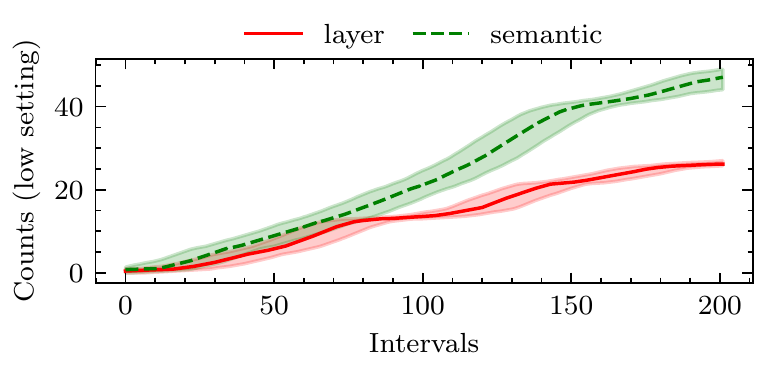}
    \label{fig:mab_counts_low}
    }\\
    \subfigure[Training Parameters with time]{
    \includegraphics[width=.31\textwidth]{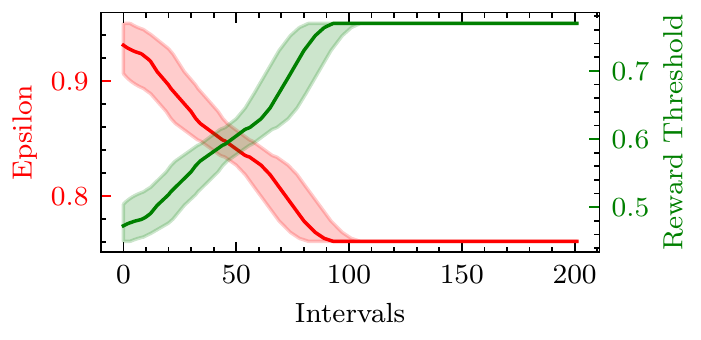}
    \label{fig:mab_params}
    }
    \subfigure[Average Rewards (SLA $>$ estimate)]{
    \includegraphics[width=.31\textwidth]{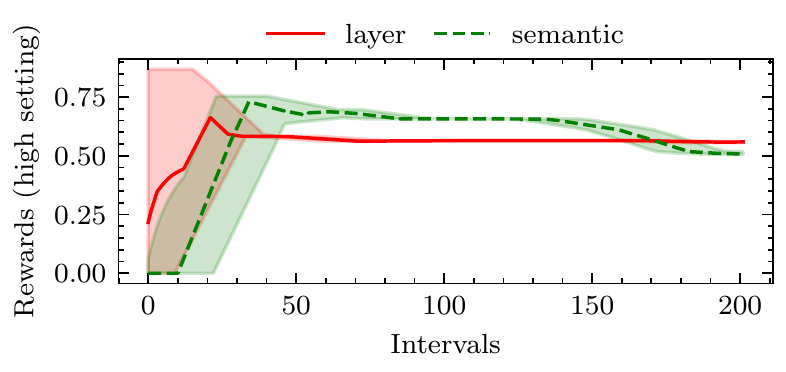}
    \label{fig:mab_rewards_high}
    }
    \subfigure[Average Rewards (SLA $<$ estimate)]{
    \includegraphics[width=.31\textwidth]{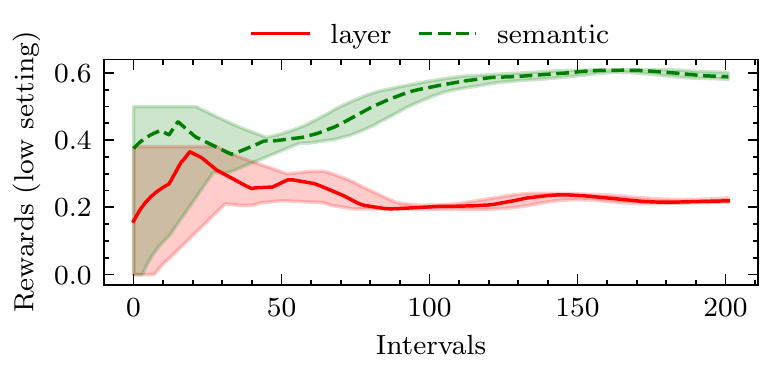}
    \label{fig:mab_rewards_low}
    }
    \caption{MAB training curves}
    \label{fig:mab}
\end{figure*}

\section{Performance Evaluation}
\label{sec:experiment}
To test the efficacy of the SplitPlace approach and compare it against the baseline methods, we perform experiments on a heterogeneous edge computing testbed. To do this we emulate a setting with mobile edge devices mounted on self-driving cars, that execute various image-recognition tasks.

\subsection{Experiment Setup}
\label{sec:setup}
As in prior work~\cite{tuli2021cosco, basu2019learn, tuli2020dynamic}, we use $\alpha = \beta = 0.5$ in \eqref{eq:reward_placement} for our experiments (we consider other value pairs in Appendix~\ref{app:ab}). Also, we use the exploration factor $c = 0.5$ for the UCB exploration and the exponential moving average parameter $\phi = 0.9$, chosen using grid-search using the cumulative reward as the metric to maximize. We create a testbed of 50 resource-constrained VMs located in the same geographical location of London, United Kingdom using Microsoft Azure. The \blue{worker} resources are shown in Table~\ref{tab:hosts}. All machines use Intel i3 2.4 GHz  processor cores with processing capacity of no more than a Raspberry Pi 4B device. To keep storage costs consistent, we keep Azure P15 Managed disk with 125 MB/s disk throughput and 256 GB size\footnote{Azure Managed Disks \url{https://docs.microsoft.com/en-us/azure/virtual-machines/disks-types\#premium-ssd}.}. The worker nodes have 4-8 GB of RAM, whereas the broker has 16 GB RAM. To factor in the mobility of the edge nodes, we use the \texttt{NetLimiter} tool to tweak the communication latency with the broker node using the mobility model described in~\cite{gilly2020modelling}. Specifically, we use the latency and bandwidth parameters of \blue{workers} from the traces generated using the Simulation of Urban Mobility (SUMO) tool~\cite{krajzewicz2012recent} that emulates mobile vehicles in a city like environment. SUMO gives us the parameters like \texttt{ping} time and network bandwidth to simulate in our testbed using \texttt{NetLimiter}. The moving averages and periodic fine-tuning allow our approach to be robust towards any kind of dynamism in the edge environment, including the one arising from mobility of worker nodes.

Our Azure environment is such that all devices are in the same LAN with 10 MBps network interface cards to avoid network bottlenecks while transferring inputs, outputs and intermediate results across neural network splits. Even if the nodes are in the same LAN with high bandwidth connections, the SUMO model would emulate the affects of mobility as is common in prior work on mobile edge computing~\cite{tuli2020dynamic, wang2022integrated}.  Further, we use the \texttt{cPickle}\footnote{\texttt{cPickle} module \url{https://docs.python.org/2/library/pickle.html\#module-cPickle}.} Python module to save the intermediate results using \texttt{bzip2} compression and \texttt{rsync}\footnote{\texttt{rsync} tool \url{https://linux.die.net/man/1/rsync}.} file-transfer utility to minimize the communication latency. For containers corresponding to a layer-split workload that are deployed in different nodes, the intermediate results are forwarded using the \texttt{scp} utility to the next container in the neural network pipeline. Similarly, for semantic splitting, the \texttt{cPickle} outputs are collected using \texttt{rsync} and concatenated using the \texttt{torch.cat} function.

We use the Microsoft Azure pricing calculator to obtain the cost of execution per hour (in US Dollars)\footnote{Microsoft Azure pricing calculator for South UK \url{https://azure.microsoft.com/en-gb/pricing/calculator/}.}. The power consumption models are taken from the Standard Performance Evaluation Corporation (SPEC) benchmarks repository\footnote{SPEC benchmark repository \url{https://www.spec.org/cloud\_iaas2018/results/}.}. The Million-Instruction-per-Second (MIPS) of all VMs are computed using the \texttt{perf-stat}\footnote{\texttt{perf-stat} tool \url{https://man7.org/linux/man-pages/man1/perf-stat.1.html}.} tool on the SPEC benchmarks. We run all experiments for 100 scheduling intervals, \textit{i.e.}, $\Gamma = 100$, with each interval being 300 seconds long, giving a total experiment time of 8 hours 20 minutes. We average over five runs and use diverse workload types to ensure statistical significance in our experiments. We consider variations of the experimental setup in Appendix~\ref{app:env}.

\begin{figure*}[]
    \centering
    \subfigure[Average Accuracy]{
    \includegraphics[width=.23\textwidth]{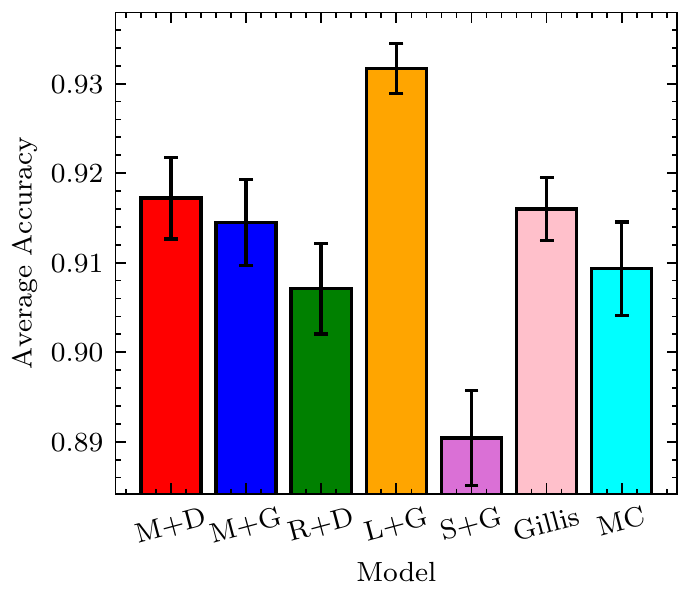}
    \label{fig:f_accuracy}
    }
    \subfigure[Average Response Time]{
    \includegraphics[width=.23\textwidth]{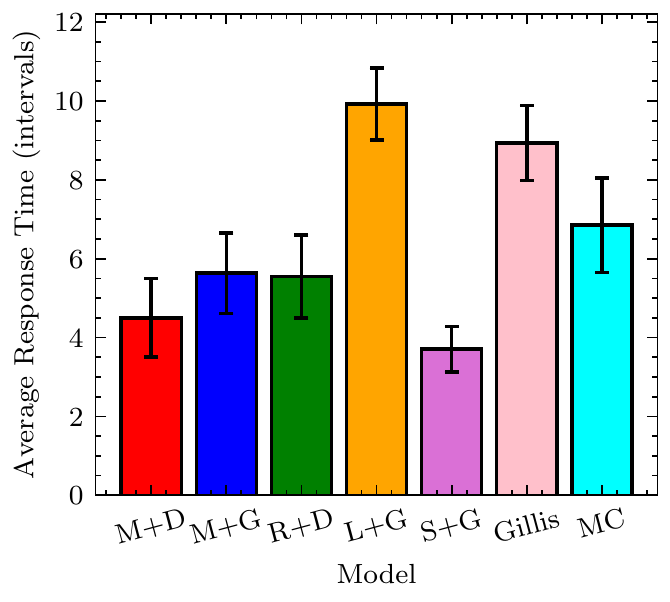}
    \label{fig:f_response}
    }
    \subfigure[Fraction of SLA Violations]{
    \includegraphics[width=.23\textwidth]{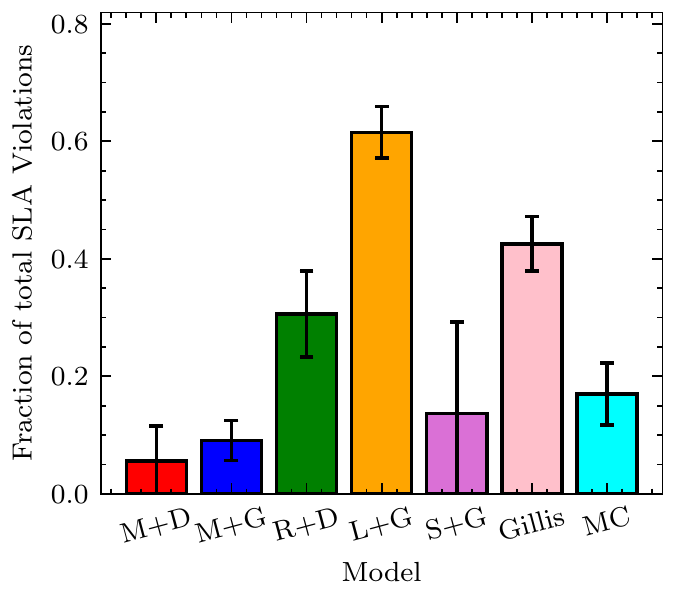}
    \label{fig:f_sla}
    }
    \subfigure[Average Reward]{
    \includegraphics[width=.23\textwidth]{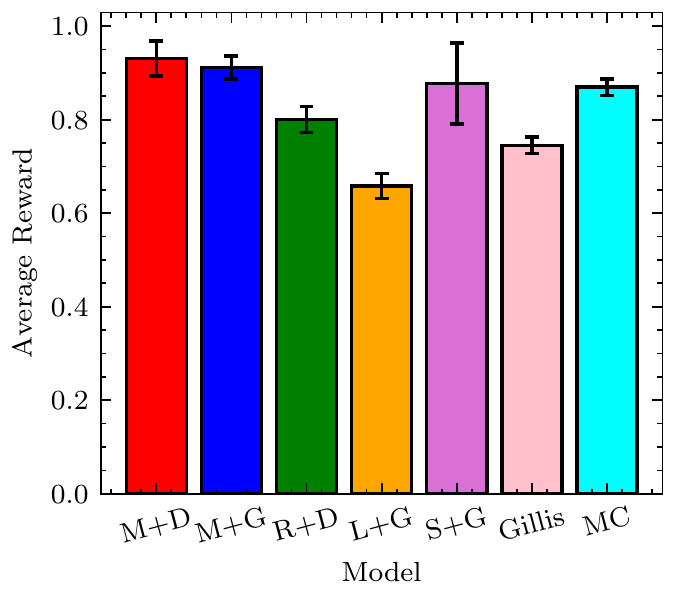}
    \label{fig:f_reward}
    }\\
    \includegraphics[width=.4\textwidth]{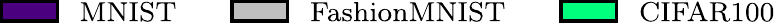}\\
    \subfigure[Average Accuracy (per application)]{
    \includegraphics[width=.23\textwidth]{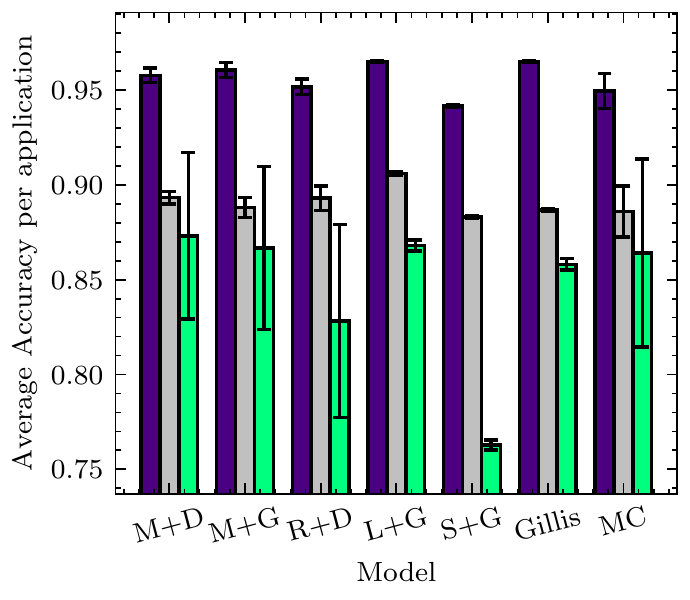}
    \label{fig:f_cpu}
    }
    \subfigure[Average Response Time (per application)]{
    \includegraphics[width=.23\textwidth]{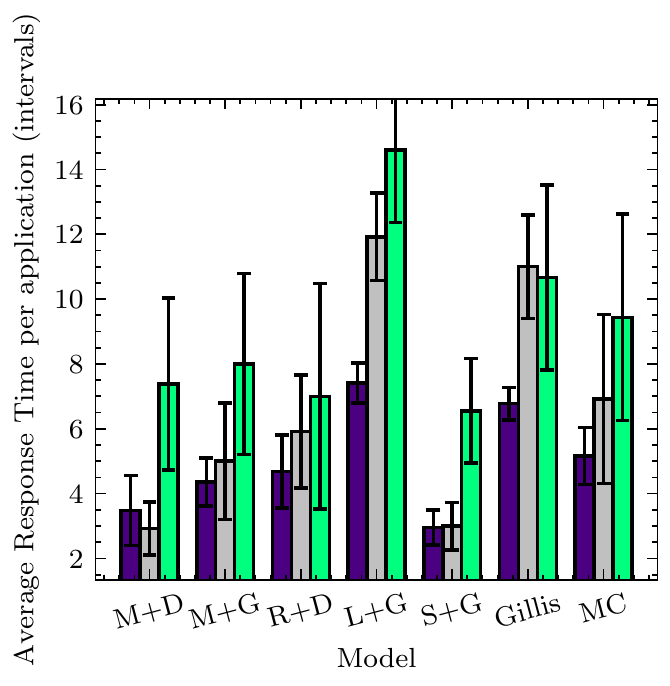}
    \label{fig:f_response_pa}
    }
    \subfigure[Fraction of SLA Violations (per application)]{
    \includegraphics[width=.23\textwidth]{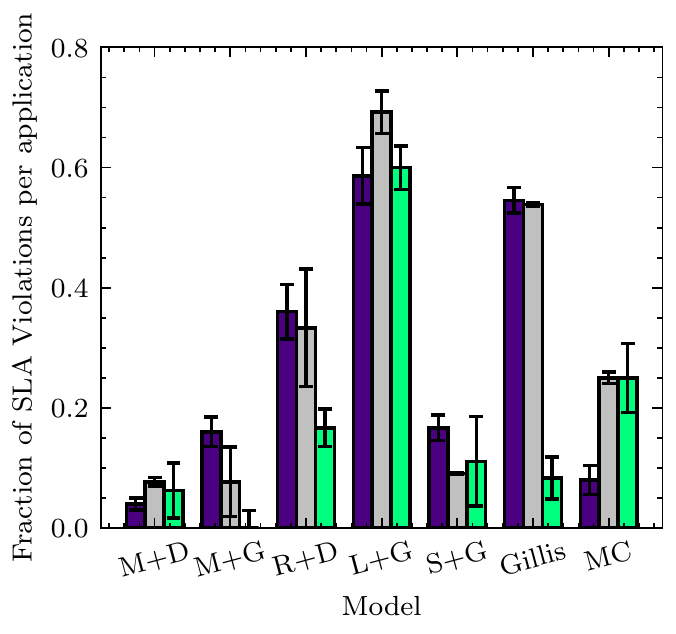}
    \label{fig:f_sla_pa}
    }
    \subfigure[Average Reward (per application)]{
    \includegraphics[width=.23\textwidth]{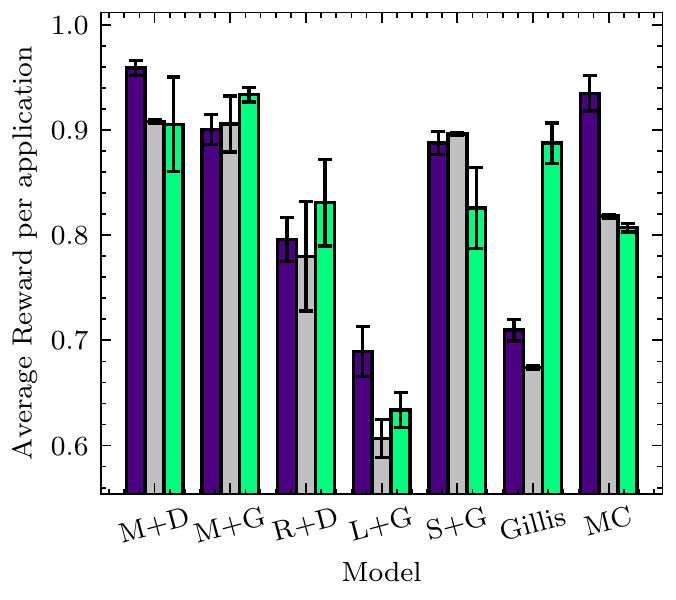}
    \label{fig:f_reward_pa}
    }
    \caption{Comparison of SplitPlace against baselines and ablated models on physical setup with 50 edge workers}
    \label{fig:framework_results}
\end{figure*}

\begin{figure*}
    \centering \setlength{\belowcaptionskip}{-8pt}
    \subfigure[Average Energy Consumption]{
    \includegraphics[width=.23\textwidth]{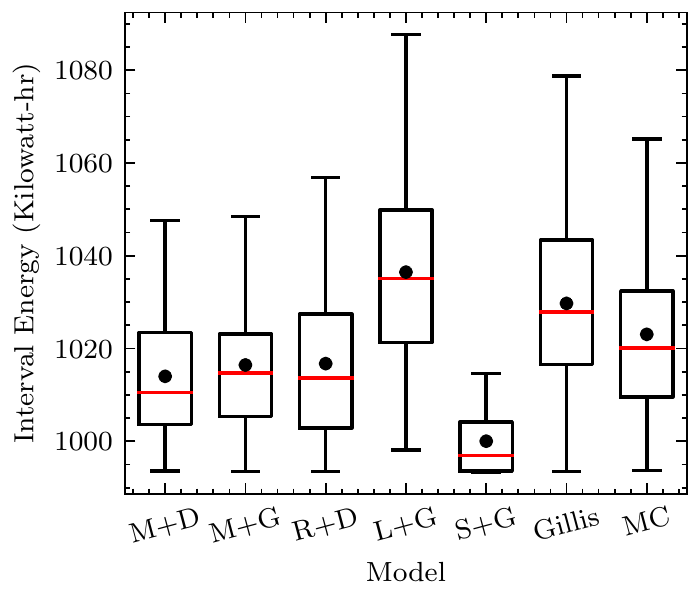}
    \label{fig:f_energy}
    }
    \subfigure[Average Execution Time]{
    \includegraphics[width=.23\textwidth]{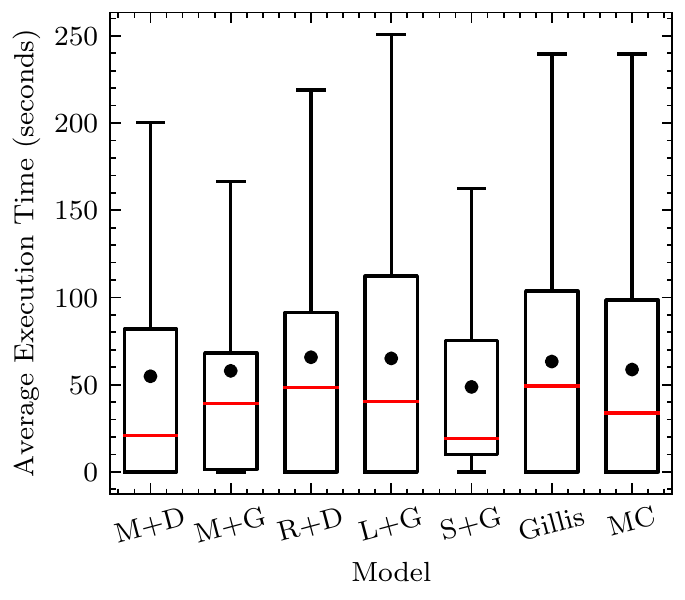}
    \label{fig:f_execution}
    }
    \subfigure[Fairness]{
    \includegraphics[width=.23\textwidth]{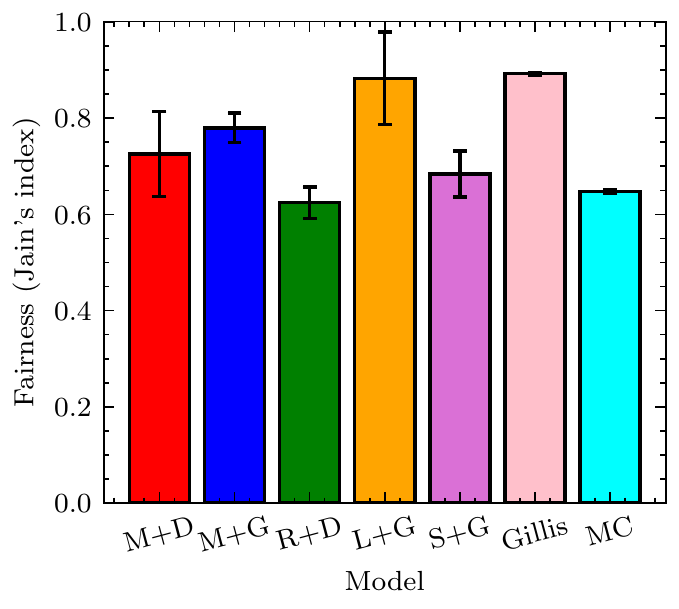}
    \label{fig:f_fairness}
    }
    \subfigure[Average Wait Time]{
    \includegraphics[width=.23\textwidth]{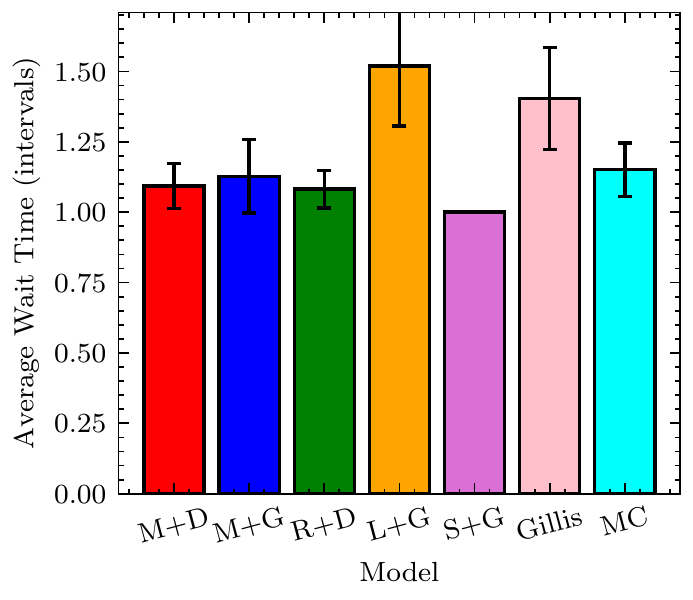}
    \label{fig:f_wait}
    }\\
    \includegraphics[width=.4\textwidth]{images/legend_app2.pdf}\\
    \subfigure[Average CPU Utilization]{
    \includegraphics[width=.23\textwidth]{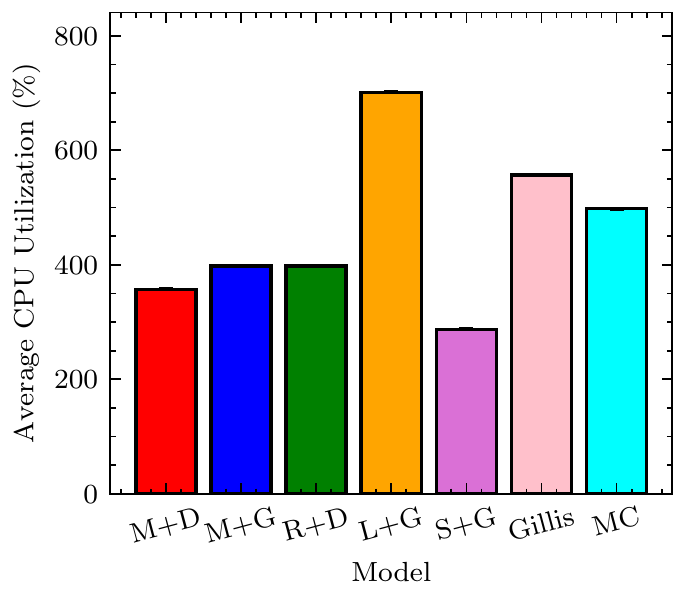}
    \label{fig:f_cpu}
    }
    \subfigure[Average RAM Utilization]{
    \includegraphics[width=.23\textwidth]{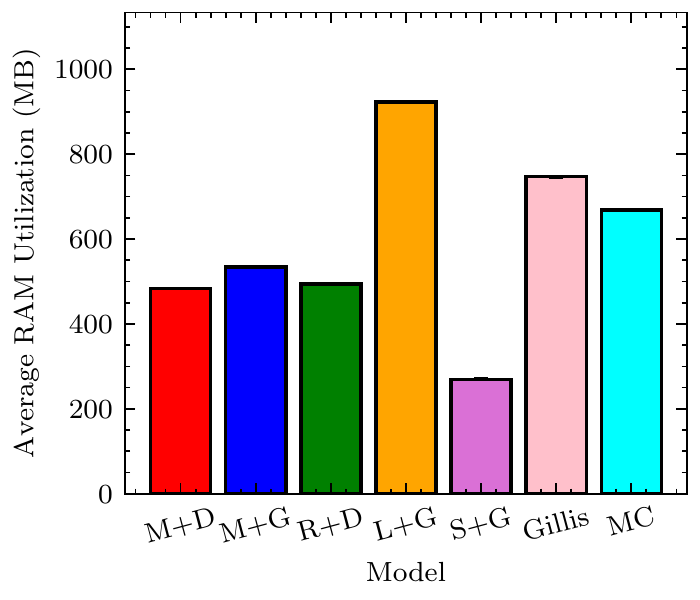}
    \label{fig:f_ram}
    }
    \subfigure[Average Fairness (per application)]{
    \includegraphics[width=.23\textwidth]{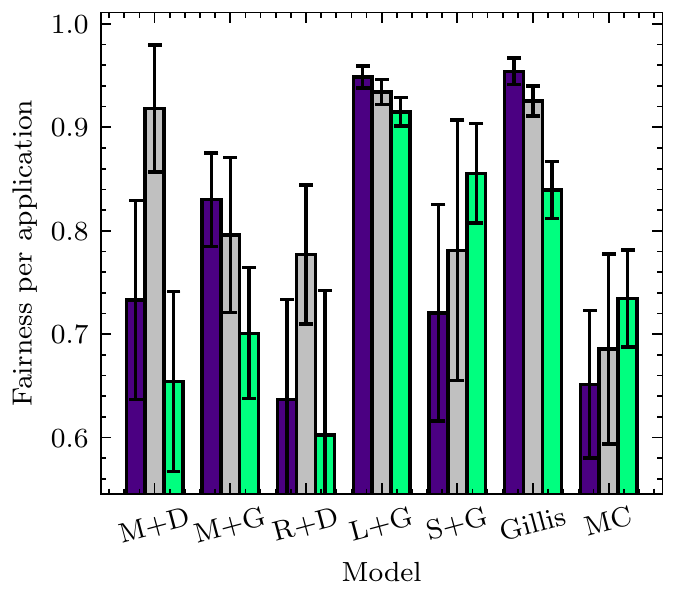}
    \label{fig:f_response_pa}
    }
    \subfigure[Average Wait Time (per application)]{
    \includegraphics[width=.23\textwidth]{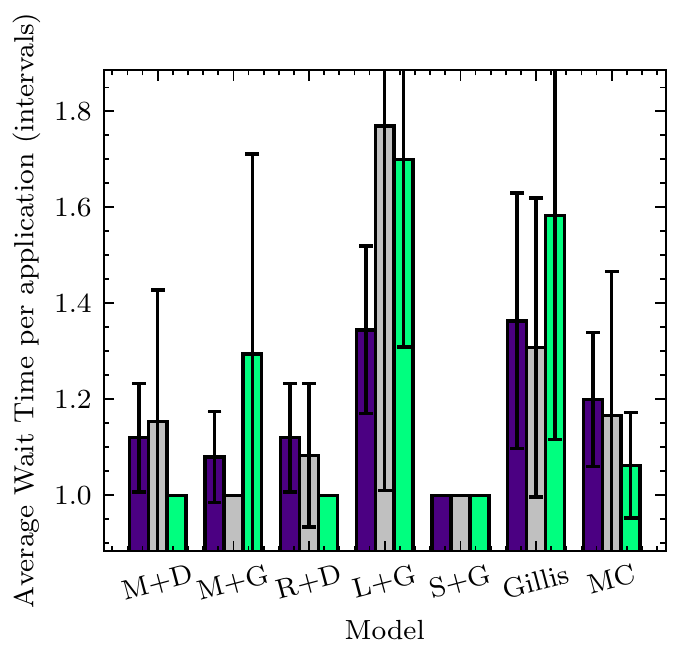}
    \label{fig:f_sla_pa}
    }\\
    \includegraphics[width=.75\textwidth]{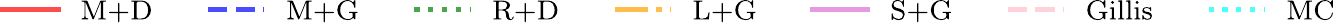}\\
    \subfigure[Average Cost per container]{
    \includegraphics[width=.23\textwidth]{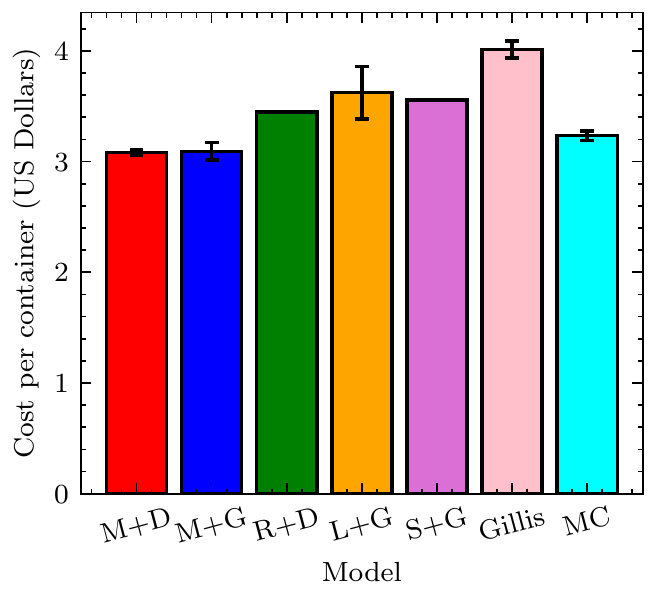}
    \label{fig:f_cost}
    }
    \subfigure[Fraction of Layer Decisions]{
    \includegraphics[width=.23\textwidth]{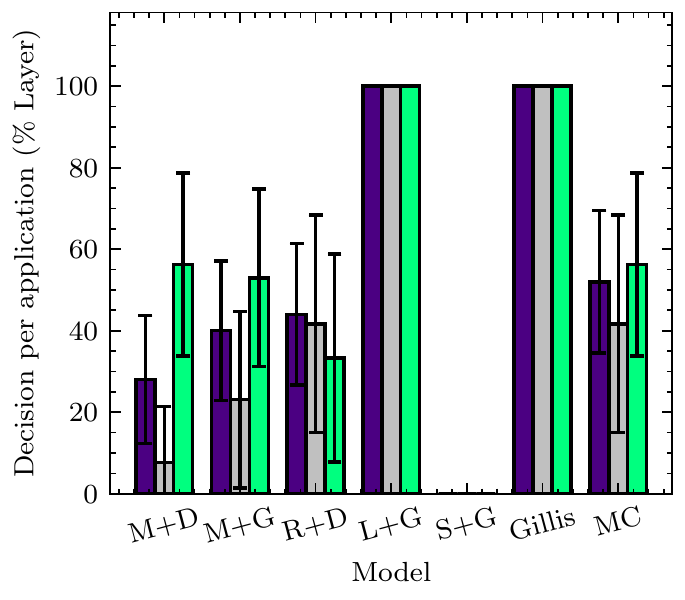}
    \label{fig:f_decision}
    }
    \subfigure[Average Accuracy with intervals]{
    \includegraphics[width=.23\textwidth]{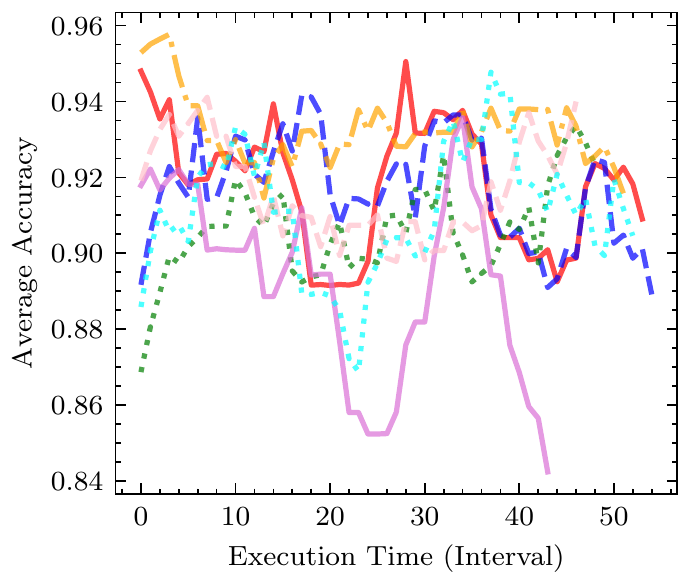}
    \label{fig:f_wait_time_series}
    }
    \subfigure[Average Response Time with intervals]{
    \includegraphics[width=.23\textwidth]{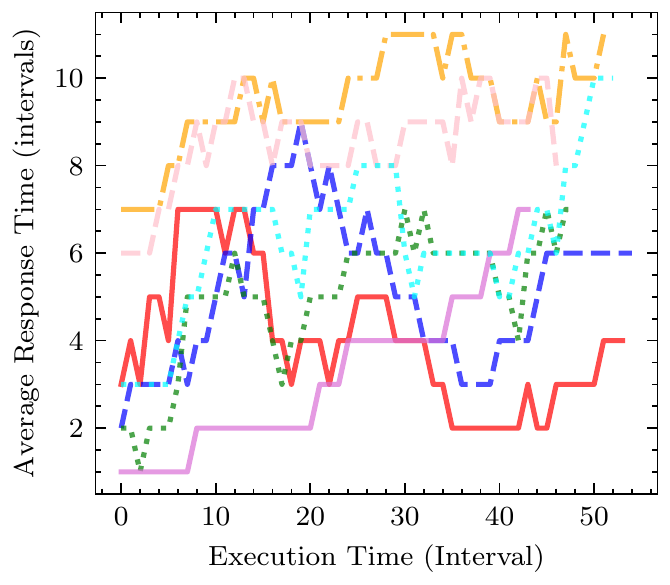}
    \label{fig:f_containers_series}
    }
    \caption{Additional results comparing SplitPlace with baselines and ablated models.}
    \label{fig:framework_results2}
\end{figure*}

\subsection{Workloads}
\label{sec:workload_setup}
Motivated from prior work~\cite{gillis}, we use three families of popular DNNs as the benchmarking models: ResNet50-V2~\cite{he2016deep}, MobileNetV2~\cite{sandler2018mobilenetv2} and InceptionV3~\cite{szegedy2017inception}. Each family has many variants of the model, each having a different number of layers in the neural model. For instance, the ResNet model has 34 and 50 layers. We use three image-classification data sets: MNIST, FashionMNIST and CIFAR100~\cite{lecun1998gradient, xiao2017fashion, krizhevsky2009learning}. MNIST is a hand-written digit recognition dataset with $28\times 28$ gray-scale images to 10-dimensional output. FashionMNIST has $28\times 28$ RGB images with 10 dimensional output. CIFAR100 has $32\times 32$ RGB images with 100-dimensional output. Thus the application set $\mathcal{A}$ becomes \{MNIST, FashionMNIST, CIFAR100\}. These models have been taken directly from the AIoTBench workloads~\cite{luo2018aiot}. This is a popular suite of AI benchmark applications for IoT and Edge computing solutions. The three specific datasets used in our experiments are motivated from the vertical use case of self-driving cars, which requires DNN-based applications to continuously recognize images with low latency requirements. Herein, an image recognition software is deployed that reads speed signs (digit recognition, MNIST), recognizes humans (through apparel and pose~\cite{hoang2018image}, FashionMNIST), identifies other objects like cars and barriers (object detection, CIFAR100). We use the implementation of neural network splitting from prior work~\cite{gillis, kim2017splitnet}.

We use the inference deadline from the work~\cite{gillis} as our SLA. To create the input tasks, we use batch sizes sampled uniformly from $16,000 - 64,000$. At the beginning of each scheduling interval, we create $Poisson(\lambda)$ tasks with $\lambda = 6$ tasks for our setup, sampled uniformly from one of the three applications~\cite{tuli2021cosco}. We consider other $\lambda$ values and single workload type (from MNIST, FashionMNIST and CIFAR100) in Appendices~\ref{app:lambda} and \ref{app:workload}. \blue{The split fragments for MNIST, FashionMNIST and CIFAR100 lead to container images of sizes 8-14 MB, 34-56 MB and 47-76 MB, respectively.} To calculate the inference accuracy to feed in the MAB models and perform UCB exploration, we also share the ground-truth labels of all datasets with all worker nodes at the time of sharing the neural models as Docker container images. We also compare edge and cloud setups in Appendix~\ref{app:cloudai} to establish the need for edge devices for latency critical workloads.

\subsection{MAB Training}

To train our MAB models, we execute the workloads on the test setup for 200 intervals and use feedback-based $\epsilon$-greedy exploration to update the layer-split decision response time estimates, $Q$-estimates and decision counts. Figure~\ref{fig:mab} shows the training curves for the two models.

Figure~\ref{fig:mab_response} shows how the response time estimates for the layer-split decision are learned starting from zero using moving averages. Figure~\ref{fig:mab_params} shows how the reward-threshold $\rho$ and decay parameter $\epsilon$ change with time. We use the decay and increment multipliers as $0.9$ and $1.1$ ($k = 0.1$ in~\eqref{eq:decay_increment}) for $\epsilon$ and $\rho$ respectively, as done in~\cite{maroti2019rbed}. Figures~\ref{fig:mab_counts_high} and~\ref{fig:mab_counts_low} show the decision counts for high and low SLA settings for both decisions. Figures~\ref{fig:mab_rewards_high} and \ref{fig:mab_rewards_low} show the $Q$-estimates for high and low SLA settings. The dichotomy between the two settings is reflected here. When the $sla_i$ of the input task $i$ is less than the estimate $R^{a_i}$ (low setting) there is a clear distinction between the rewards of the two decisions as layer-split is likely to lead to SLA violation and hence lower rewards. However, when $sla_i$ is greater than the estimate $R^{a_i}$ (high setting), both decisions give relatively high rewards with layer-split decision slightly surpassing the semantic-split due to higher average accuracy as discussed in Section~\ref{sec:motivation}.

The feedback-based $\epsilon$-greedy training allows us to obtain close estimates of the average response times of the layer-split executions for each application type and average rewards for both decisions in high and low SLA settings. Thus, in our experiments, we initialize the expected reward ($Q$) and layer-split response time ($R$) estimates by the values we get from this training approach. At test time, we dynamically update these estimates using~\eqref{eq:r_update} and~\eqref{eq:q_update}.

\subsection{Performance Metrics}

We use the following evaluation metrics in our experiments as motivated from prior works~\cite{gill2019transformative, tuli2021cosco, basu2019learn}. We also use $AEC$ and $ART$ as discussed in Section~\ref{sec:splitplace}.

\begin{enumerate}[leftmargin=*]
    \item \textit{Average Accuracy} is defined for an execution trace as the average accuracy of all tasks run in an experiment, \textit{i.e},
    \begin{equation}
    \label{eq:accuracy}
        \begin{aligned}
        Accuracy = \frac{\sum_t \sum_{i \in E_t} p_i}{\sum_t |E_t|}.
        \end{aligned}
    \end{equation}
    \item \textit{Fraction of SLA violation} is defined for an execution trace as the fraction of all tasks run in an experiment for which the response time is higher than the SLA deadline, \textit{i.e.},
    \begin{equation}
    \label{eq:sla}
        \begin{aligned}
        SLA\ Violations = \frac{\sum_t \sum_{t \in E_t} \mathbbm{1}(sla_i \geq r_i)}{\sum_t |E_t|}.
        \end{aligned}
    \end{equation}
    \item \textit{Average Reward} is defined for an execution trace as follows
    \begin{equation}
    \label{eq:avg_reward}
        \begin{aligned}
        Reward =  \frac{\sum_t \sum_{t \in E_t} \mathbbm{1}(sla_i \geq r_i) + p_i}{2 \cdot \sum_t |E_t|}.
        \end{aligned}
    \end{equation}
    \item \textit{Execution Cost} is defined for an execution trace as the total cost incurred during the experiment, \textit{i.e.},
    \begin{equation}
    \label{eq:cost}
        \begin{aligned}
        Cost =  \sum_{h\in \mathcal{H}} \int_{x} C_h(x)dx.
        \end{aligned}
    \end{equation}
    where $C_h(x)$ is the cost function for \blue{worker} $h$ with time.
    \item \textit{Average Wait Time} is the average time a task had to wait in the wait queue till it could be allocated to a \blue{worker} for execution.
    \item \textit{Average Execution Time} is the response time minus the wait time, averaged for all tasks run in an experiment.
    \item \textit{Fairness} is defined as the Jain's fairness index for execution on tasks over the \blue{edge workers}~\cite{tuli2021cosco}.
\end{enumerate}

\begin{table*}[t]
    \centering
    \caption{Comparison of SplitPlace with baseline and ablated models. The best achieved value for each metric is shown in bold. Units: Energy (MW-hr), Scheduling Time (seconds), Fairness (Jain's Index), Wait Time (Intervals), Response Time (Intervals).}
    \begin{tabular}{@{}lcccccccc@{}}
    \toprule 
    \multirow{2}{*}{\textbf{Model}} & \multirow{2}{*}{\textbf{Energy}} & \textbf{Scheduling} & \multirow{2}{*}{\textbf{Fairness}} & \textbf{Wait} & \textbf{Response} & \textbf{SLA} & \multirow{2}{*}{\textbf{Accuracy}} & \textbf{Average}\tabularnewline
     &  & \textbf{Time} &  & \textbf{Time} & \textbf{Time} & \textbf{Violations} &  & \textbf{Reward}\tabularnewline
    \midrule
    \multicolumn{9}{c}{\textbf{Baselines}}\tabularnewline
    \midrule 
    Model Compression & 1.1368  & 8.84$\pm$0.02 & 0.65$\pm$0.01 & 1.15$\pm$0.09 & 6.85$\pm$1.20 & 0.26$\pm$0.02 & 89.93 & 83.98\tabularnewline

    Gillis & 1.1442  & \textbf{8.22$\pm$0.01} & \textbf{0.89$\pm$0.03} & 1.40$\pm$0.18 & 8.39$\pm$0.95 & 0.22$\pm$0.03 & 91.90 & 84.17\tabularnewline
    \midrule 
    \multicolumn{9}{c}{\textbf{Ablation}}\tabularnewline
    \midrule 
    Semantic+GOBI & 1.1112  & 8.68$\pm$0.03 & 0.68$\pm$0.04 & 1.08$\pm$0.00 & \textbf{3.70$\pm$0.57} & 0.14$\pm$0.04 & 89.04 & 83.91\tabularnewline

    Layer+GOBI & 1.1517  & 8.72$\pm$0.01 & 0.88$\pm$0.03 & 1.52$\pm$0.21 & 9.92$\pm$0.91 & 0.62$\pm$0.07 & \textbf{93.17} & 64.87\tabularnewline

    Random+DASO & 1.1297  & 8.86$\pm$0.01 & 0.62$\pm$0.05 & \textbf{1.00$\pm$0.07} & 5.55$\pm$1.05 & 0.29$\pm$0.09 & 90.71 & 81.62\tabularnewline

    MAB+GOBI & 1.1290  & 9.12$\pm$0.02 & 0.78$\pm$0.08 & 1.13$\pm$0.13 & 5.64$\pm$1.02 & 0.10$\pm$0.03 & 91.45 & 90.18\tabularnewline
    \midrule 
    \multicolumn{9}{c}{\textbf{SplitPlace Model}}\tabularnewline
    \midrule 
    MAB+DASO & \textbf{1.0867 } & 9.32$\pm$0.02 & 0.73$\pm$0.01 & 1.09$\pm$0.08 & 4.50$\pm$1.00 & \textbf{0.08$\pm$0.02} & 92.72 & \textbf{94.18}\tabularnewline
    \bottomrule 
    \end{tabular}
    \label{tab:framework}
\end{table*}

\subsection{Baselines and Ablated Models}

We compare the performance of the SplitPlace approach against the state-of-the-art baselines \textit{Gillis} and BottleNet++ Model Compression (denoted as \textit{MC} in our graphs)~\cite{gillis, eshratifar2019bottlenet, shao2020bottlenet++}. Gillis refers to the reinforcement learning method proposed in~\cite{gillis} that leverages both layer-splitting and compression models to achieve optimal response time and inference accuracy. Note that contrary to the original Gillis' work, our implementation does not leverage serverless functions. MC is a model-compression approach motivated from BottleNet++ that we implement using the \texttt{PyTorch Prune} library.\footnote{PyTorch Prune. \url{https://pytorch.org/docs/stable/generated/torch.nn.utils.prune.ln_structured.html}. Accessed 10 October 2021.} Further details in Section~\ref{sec:related}. We do not include results for other methods discussed in Section~\ref{sec:related} as MC and Gillis give better results empirically for all comparison metrics. We also compare SplitPlace with ablated models, where we replace one or both of the MAB or DASO components with simpler versions as described below.

\begin{itemize}
    \item \textit{Semantic+GOBI (S+G)}: Semantic-split decision only with vanilla GOBI placement module.
    \item \textit{Layer+GOBI (L+G)}: Layer-split decision only with vanilla GOBI placement module.
    \item \textit{Random+DASO (R+D)}: Random split decision with DASO placement module.
    \item \textit{MAB+GOBI (M+G)}: MAB based split decider with vanilla GOBI placement module.
\end{itemize}

The final SplitPlace approach is represented as MAB+DASO or M+D in shorthand notation in the graphs. These ablated baselines help us determine the relative improvements in performance by the two components of MAB and DASO separately.

\subsection{Results and Ablation Analysis}

We now provide comparative results showing the performance of the proposed SplitPlace approach against the baseline models and argue the importance of the MAB and decision-aware placement using ablation analysis. We train the \textit{GOBI} and \textit{DASO} models using the execution trace dataset used to train the MAB models. The learning rate ($\eta$) was set to $10^{-3}$ from~\cite{tuli2021cosco}.

Figure~\ref{fig:framework_results} shows the average reward and related performance metrics, \textit{i.e.}, accuracy, response time and SLA violation rate. As expected, the L+G policy gives the highest accuracy of $93.17\%$ as all decisions are layer-wise only with a higher inference performance than semantic-split execution. The S+G policy gives the least accuracy of $89.04\%$. However, due to layer-splits only the L+G policy also has the highest average response time, subsequently giving the highest SLA violation rate. On the other hand, S+G policy has the least average response time. However, due to the intelligent decision making in SplitPlace, it is able to get the highest total reward of $0.9418$. Similar trends are also seen when comparing across models for each application. The accuracy is the highest for the MNIST dataset and lowest for CIFAR100. Average response time is highest for the CIFAR100 and lowest for MNIST in general. Among the baselines, the Gillis approach has the lowest SLA violation rate of $22\%$ and SplitPlace improves upon this by giving $14\%$ lower SLA violations (only $8\%$). Gillis has higher accuracy between the baselines of $91.9\%$, with SplitPlace giving an average improvement of $0.82\%$. Overall, the total reward of SplitPlace is higher than the baselines by at least $10.1\%$, giving the reward of $94.18\%$. 

Figure~\ref{fig:framework_results2} shows the performance of all models for other evaluation metrics like energy, execution time and fairness. Compared to the baselines, SplitPlace can reduce energy consumption by up to $4.41\% - 5.03\%$ giving an average energy consumption of $1.0867$ MW-hr. However, the SplitPlace approach has higher scheduling time and lower fairness index (Table~\ref{tab:framework}). The Gillis baseline has the highest fairness index of $0.89$, however this index for SplitPlace is $0.73$. SplitPlace has a higher overhead of $11.8\%$ compared to the Gillis baseline in terms of scheduling time. Figure~\ref{fig:f_cost} compares the average execution cost (in USD) for all models. As SplitPlace is able to run the maximum number of containers in the 100 intervals, it has the least cost of $3.07$ USD/container.  The main advantage of SplitPlace is the intelligent splitting decisions facilitate overcoming the memory bottlenecks in edge environments, giving up to 32\% lower RAM utilization compared to Gillis and Model Compression.

\blue{In terms of the initial communication time of the Docker container images, the SplitPlace method takes 30 seconds at the start of an execution. Gillis and MC have such communication times of 20 and 18 seconds, respectively. This demonstrates that SplitPlace has a low one-time overhead (up to 12 seconds) compared to the baselines when compared to the gains in response time (up to 46\%) that linearly scales as the number of workloads increase.}

A summary of comparisons with values of main performance metrics for all models is given in Table~\ref{tab:framework}. The best values achieved for each metric are highlighted in bold.

\section{Conclusions}
\label{sec:conclusion}

In this work, we present SplitPlace, a novel framework for efficiently managing demanding neural network based applications. SplitPlace exploits the trade-off between layer and semantic split models where the former gives higher accuracy, but the latter gives much lower response times. This allows SplitPlace to not only manage tasks to maintain high inference accuracy on average, but also reduce SLA violation rate. The proposed model uses a Multi-Armed-Bandits based policy to decide which split strategy to use according to the SLA deadline of the incoming task. Moreover, it uses a decision-aware learning model to take appropriate placement decisions for those neural fragments on mobile edge \blue{workers}. Further, both MAB and learning models are dynamically tuned to adapt to volatile scenarios. All these contributions allow SplitPlace to out-perform the baseline models in terms of average response time, SLA violation rate, inference accuracy and total reward by up to $46.3\%$, $69.2\%$, $3.1\%$ and $12.1\%$ respectively in a heterogeneous edge environment with real-world workloads. 

We propose the following future directions for this work. An extension of the current work may be developed that dynamically updates the splitting configuration to adapt to more heterogeneous and non-stationary edge environments~\cite{bonawitz2019towards}. Moreover, the current model assumes that all neural models are divisible into independent layers. This may be hard for deep learning models like attention based neural networks or transformer models~\cite{vaswani2017attention}. Finally, the model only considers splits and their placement as containers, more fine-grained methods involving Neural Architecture Search and cost efficient deployment methods may be explored like serverless frameworks~\cite{casale2020radon}. \blue{Other considerations such as privacy concerns and non-stationary number of active edge nodes with extreme levels of heterogeneity such that the placement decision has a significant impact on response time is also part of future work.}

% Accuracy critical use cases would need a hybrid splitting approach that could give higher accuracy than pure semantic splits but not as high overheads as layer splits. Further, the framework can be extended to allow distributed split nets training using federated learning for data sharing and reduced bandwidth overheads~\cite{bonawitz2019towards}. Moreover, the SplitPlace method assumes all devices are in the same LAN, allowing us to separate our method from network bandwidth limitations. However, for more geo-distributed environments, intelligent splitting would be required to prevent large data sharing overheads on the network.

\section*{Software Availability}
% The code and relevant training scripts shall be made publicly available on GitHub under BSD-3 licence upon publication.
The code is available at \url{https://github.com/imperial-qore/SplitPlace}. The Docker images used in the experiments are available at \url{https://hub.docker.com/u/shreshthtuli}.

\section*{Acknowledgments}
{Shreshth Tuli is grateful to the Imperial College London for funding his Ph.D. through the President’s Ph.D. Scholarship scheme. We thank Feng Yan for helpful discussions.}

\bibliographystyle{IEEEtran}
\bibliography{references}

\appendices

\begin{figure*}[t]
    \centering
    \includegraphics[width=.8\textwidth]{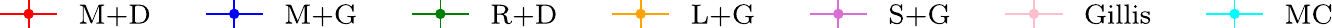}\\
    \subfigure[Average Accuracy]{
    \includegraphics[height=.221\textwidth]{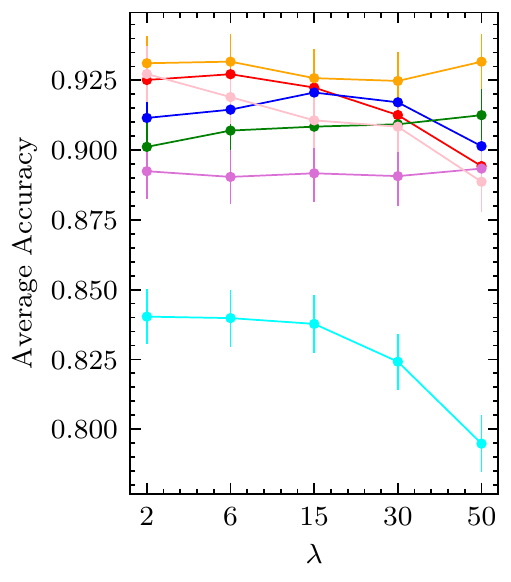}
    \label{fig:s_sensl_a}
    }
    \subfigure[Average Response Time (Intervals)]{
    \includegraphics[height=.221\textwidth]{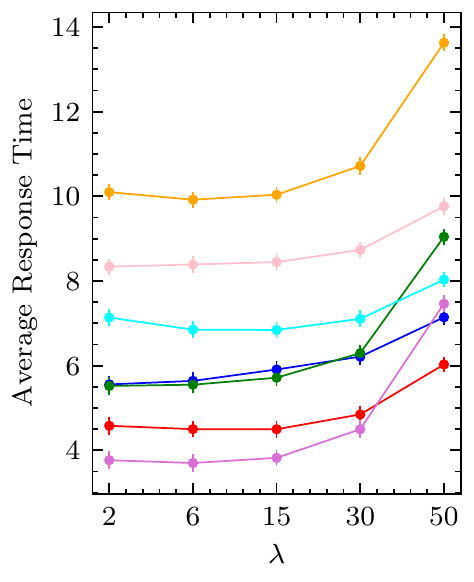}
    \label{fig:s_sensl_r}
    }
    \subfigure[Fraction of SLA Violations]{
    \includegraphics[height=.221\textwidth]{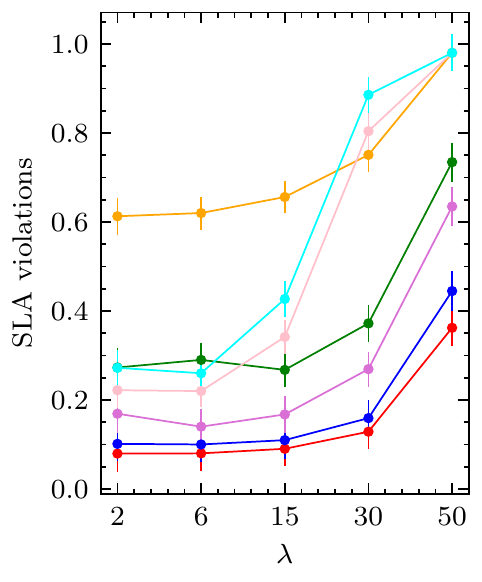}
    \label{fig:s_sensl_sla}
    }
    \subfigure[Average Reward]{
    \includegraphics[height=.221\textwidth]{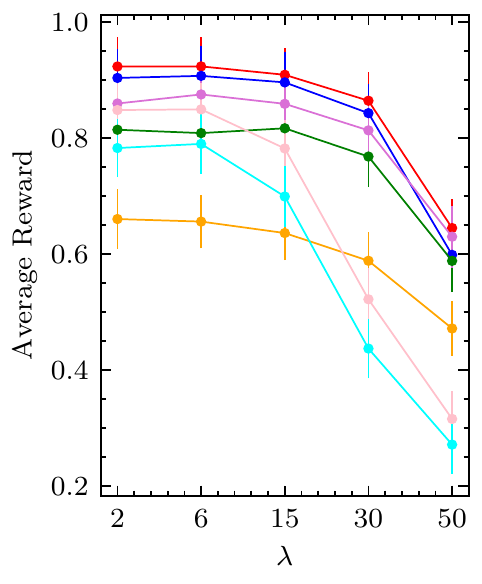}
    \label{fig:s_sensl_reward}
    }
    \subfigure[Average Energy (MW-hr)]{
    \includegraphics[height=.221\textwidth]{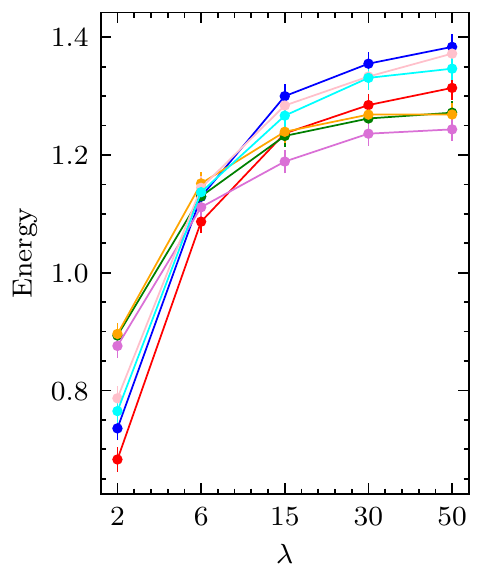}
    \label{fig:s_sensl_e}
    }
    \caption{Sensitivity of $\lambda$ on all models.}
    \label{fig:sensl}
\end{figure*}

\begin{figure*}[t]
    \centering
    \includegraphics[width=.6\textwidth]{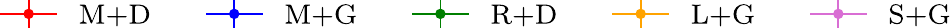}\\
    \subfigure[Average Accuracy]{
    \includegraphics[height=.222\textwidth]{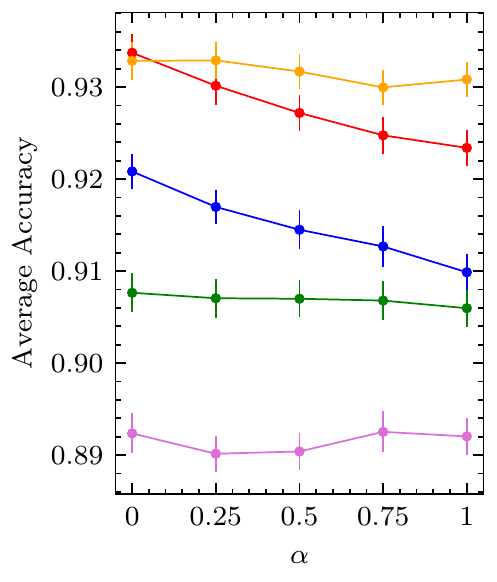}
    \label{fig:s_sensab_a}
    }
    \subfigure[Average Response Time (Intervals)]{
    \includegraphics[height=.222\textwidth]{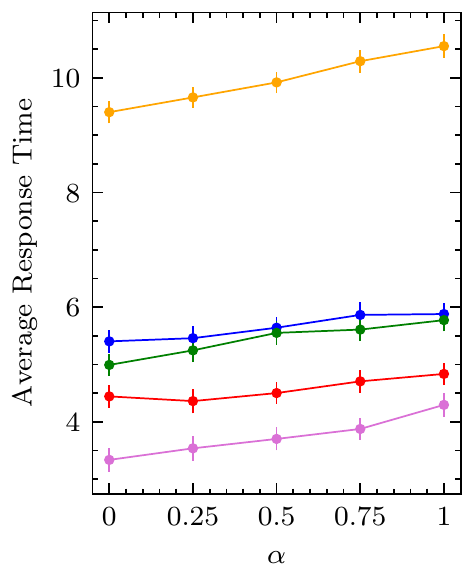}
    \label{fig:s_sensab_r}
    }
    \subfigure[Fraction of SLA Violations]{
    \includegraphics[height=.222\textwidth]{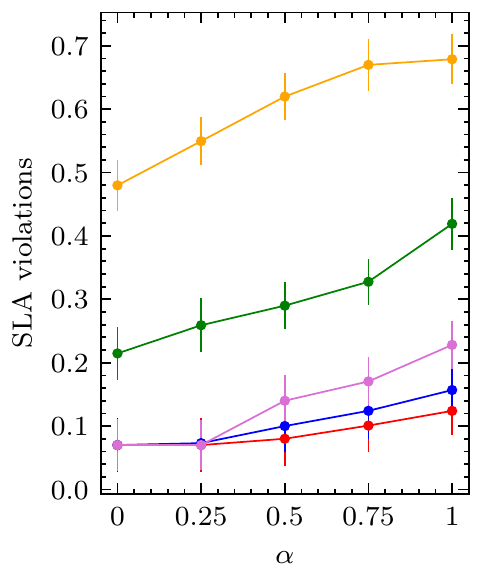}
    \label{fig:s_sensab_sla}
    }
    \subfigure[Average Reward]{
    \includegraphics[height=.222\textwidth]{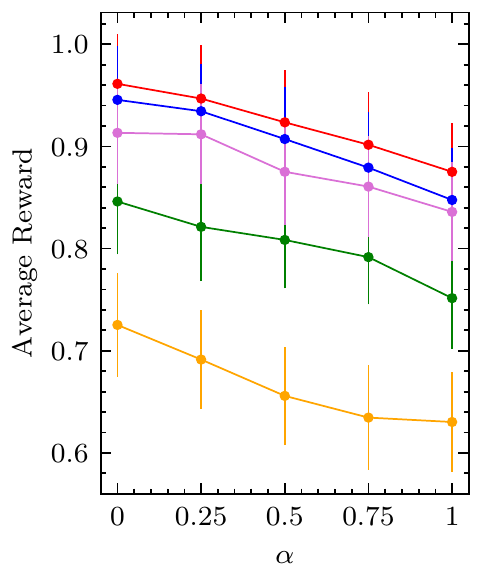}
    \label{fig:s_sensab_reward}
    }
    \subfigure[Average Energy (MW-hr)]{
    \includegraphics[height=.222\textwidth]{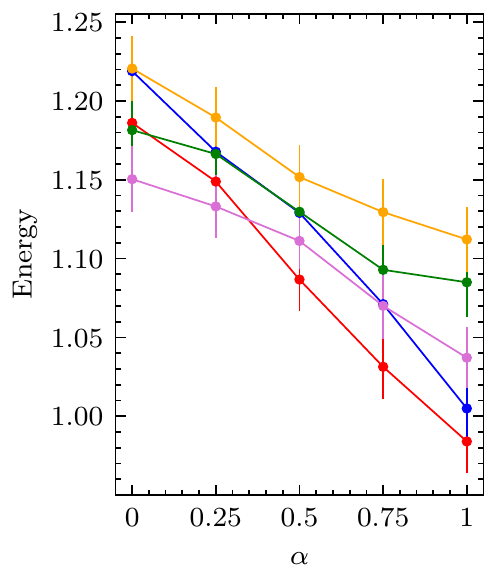}
    \label{fig:s_sensab_e}
    }
    \caption{Sensitivity of $\alpha, \beta$ on ablated SplitPlace models.}
    \label{fig:sensab}
\end{figure*}

\section{Additional Experiments}
\label{sec:appendix}

\begin{figure}
    \centering
    \includegraphics[width=0.72\linewidth]{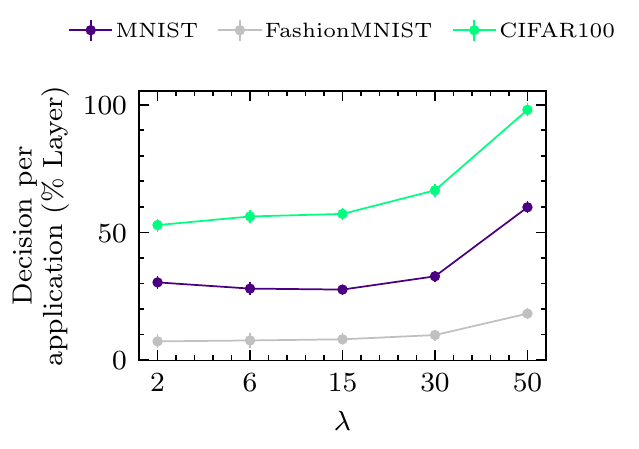}
    \caption{Fraction of layer-split decisions of the SplitPlace model with increasing $\lambda$.}
    \label{fig:dec_l}
\end{figure}
\begin{figure}
    \centering
    \includegraphics[width=0.72\linewidth]{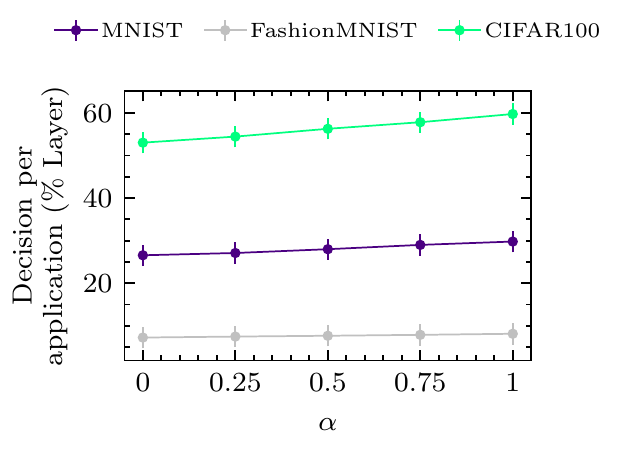}
    \caption{Fraction of layer-split decisions of the SplitPlace model with increasing $\alpha$.}
    \label{fig:dec_ab}
\end{figure}

\subsection{Sensitivity with $\lambda$}
\label{app:lambda}
We perform sensitivity analysis with the number of workloads being generated at the start of each scheduling interval to test the scalability of the SplitPlace model. As described in Section~\ref{sec:workload_setup}, we create $Poisson(\lambda)$ tasks at the beginning of each scheduling interval with $\lambda = 6$ for our setup. The tasks are sampled uniformly from the three applications: MNIST, FashionMNIST and CIFAR100 as done in our experiments in Section~\ref{sec:experiment}. We vary the $\lambda$ parameter from $2$ to $50$ to test how the performance varies with this parameter. Beyond $\lambda = 50$, the baseline models diverge and the system breaks down due to computational overload for these models. Figure~\ref{fig:sensl} demonstrates the average accuracy, response time, fraction of SLA violations, reward and energy consumption for all models with varying the $\lambda$ parameter. Again, all experiments are performed for 100 scheduling intervals with the same setup as described in Section~\ref{sec:setup}. Clearly, as $\lambda$ increases, the number of workloads and hence the resource requirements increase.  

The average accuracy observed with policies that take the splitting decision based on external reward signals (M+D, M+G, MC and Gillis) drops as the $\lambda$ parameter increases. This is not seen for policies R+D, L+G or S+G, as their splitting decision (which affects the classification accuracy) are unaffected. The other policies see a gradual decline of accuracy by $2.61\%-4.83\%$ as the ratio of semantic-split decisions increases to reduce resource consumption (see Figure~\ref{fig:dec_l}). However, we observe that the response times increase with $\lambda$. This is due to the increasing wait times and contention affects in congested edge environments. Still, the SplitPlace model (M+D) maintains a low average response time of $4.76$ intervals, $33.61\%-45.31\%$ lower than the baseline models (MC and Gillis). All models see a sharp increase in violation rate as $\lambda$ increases to 30 tasks in terms of SLA violations. The L+G, MC and Gillis models show average violation rates of nearly 1 for $\lambda = 50$. However, the average slope is least for the SplitPlace model, increasing from $8\%$ to $35.1\%$ (increase of $27\%$), whereas for others this change is much higher $34.87\%-76.46\%$. Naturally, as SLA violation rates increase and accuracies decrease, the average reward would decrease (Figure~\ref{fig:s_sensl_reward}). The average rewards of the M+D model ($88.44$) are still higher than the baselines and ablated models ($64.69-85.66$). Figure~\ref{fig:s_sensl_e} shows the change of energy consumption as $\lambda$ increases. This metric sharply increases as $\lambda$ changes from 2 tasks to 6 tasks. Upon further increase, the energy does not increases as significantly as edge environment reaches a saturation point and no more tasks can be scheduled. All models have similar trends in terms of the energy consumption metric.

\subsection{Sensitivity with $\alpha, \beta$}
\label{app:ab}

We also perform sensitivity analysis on the $\alpha, \beta$ parameters introduced in Section~\ref{sec:placement_module} for the SplitPlace and its ablated versions. $\alpha$ and $\beta$ are the weights corresponding to the average energy consumption (AEC) and average response time (ART) metrics in Equation~\ref{eq:reward_placement}. Note that it is a convex combination of the two metrics, so $\alpha+\beta = 1$. These experiments help us test the robustness of the SplitPlace model in the presence of diverse QoS objectives.

Figure~\ref{fig:sensab} shows the change in performance metrics when $\alpha$ varies from 0 to 1. Some observations overlap with Figure~\ref{fig:sensl}. In terms of energy consumption, all models show a decline in this metric as $\alpha$ increases. This is because of the energy conserving scheduling decisions by the GOBI model, allocating a higher fraction of tasks to the low-power edge nodes (B2ms and E2asv4). This leads to higher resource contention in these devices, leading to higher SLA violations, for which the MAB model tends to take a higher ratio of semantic-split decisions (due to their lower response times, see Figure~\ref{fig:dec_ab}). This leads to the decline in the accuracy of the M+D and M+G models as $\alpha$ increases, whereas the average accuracy for reward-free models (R+D, L+G, S+G) remains nearly unchanged. The average response times and SLA violation rates increase steadily for all models. However, the increase is not as significant for the MAB based models (M+D and M+G) due to the UCB based external signal. The average reward is still the highest for the SplitPlace model ($91.93$) compares to the ablated models ($65.91-90.39$).

\subsection{Constraining Environments}
\label{app:env}
We now perform the same experiments with various constraints in the original edge computing environment. These experiments demonstrate the ability of the MAB based models to adapt to varied deployment scenarios efficiently.

\textbf{Changes in Setup:} For each host in Table~\ref{tab:hosts}, we limit the core count (compute constrained), network bandwidth (network constrained) and RAM size (memory constrained) to half the original value\footnote{To limit the CPU core usage in Linux kernel we modify the \texttt{grubconfig}. To limit the RAM usage, we use the \texttt{ulimit} tool. To limit the network bandwidth, we appropriately modify the parameters of the \texttt{Netlimiter} tool.}. These setups show us the performance of the SplitPlace model with respect to the baselines techniques in more diverse experimental scenarios.

\begin{figure*}[t]
    \centering
    \includegraphics[width=.8\textwidth]{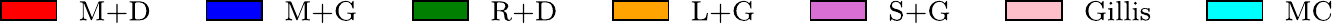}\\
    \subfigure[Average Accuracy]{
    \includegraphics[width=.236\textwidth]{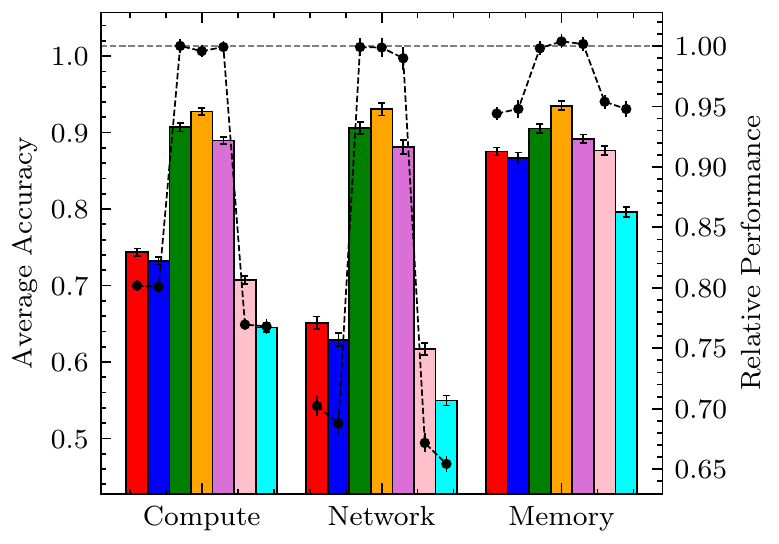}
    \label{fig:s_env_a}
    }
    \subfigure[Average Response Time]{
    \includegraphics[width=.236\textwidth]{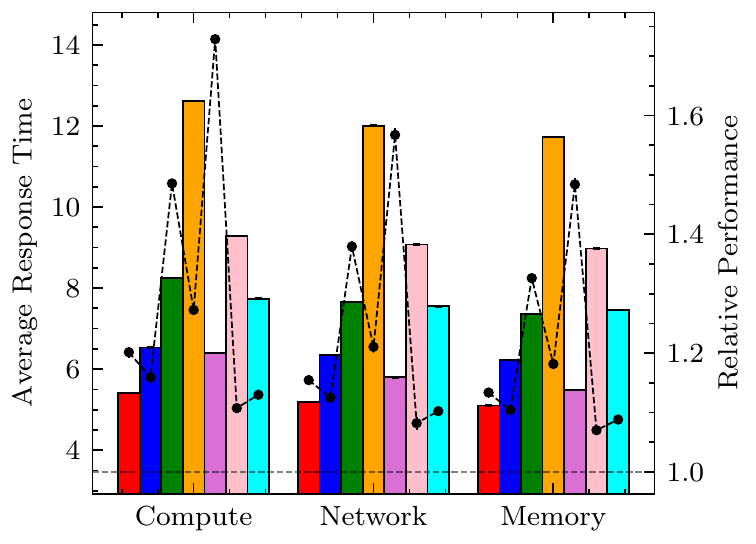}
    \label{fig:s_env_r}
    }
    \subfigure[Fraction of SLA Violations]{
    \includegraphics[width=.236\textwidth]{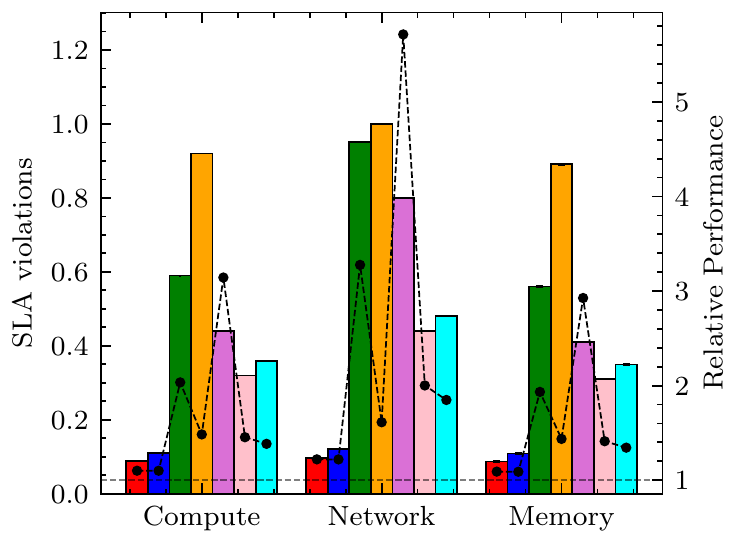}
    \label{fig:s_env_sla}
    }
    \subfigure[Average Reward]{
    \includegraphics[width=.236\textwidth]{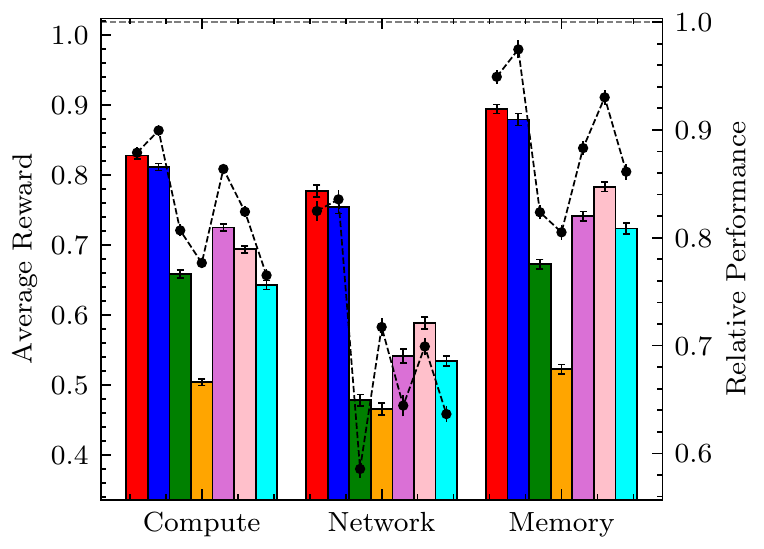}
    \label{fig:s_env_reward}
    }
    \caption{Comparison of SplitPlace against baselines and ablated models in Compute, Network and Memory constrained environments.}
    \label{fig:s_env}
\end{figure*}
\begin{figure*}[t]
    \centering
    \includegraphics[width=.9\textwidth]{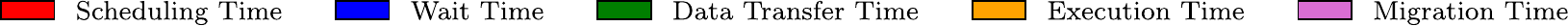}\\
    \subfigure[Normal Setup]{
    \includegraphics[width=.236\textwidth]{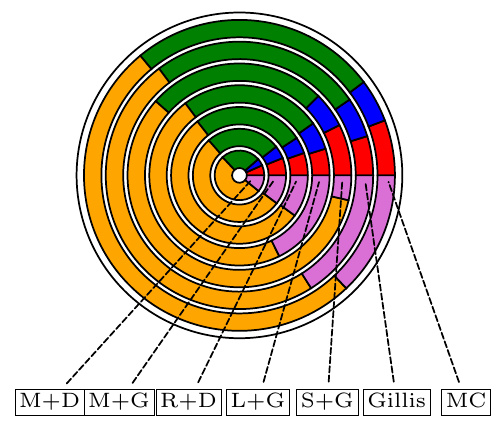}
    \label{fig:s_env_pie_normal}
    }
    \subfigure[Compute Constrained]{
    \includegraphics[width=.236\textwidth]{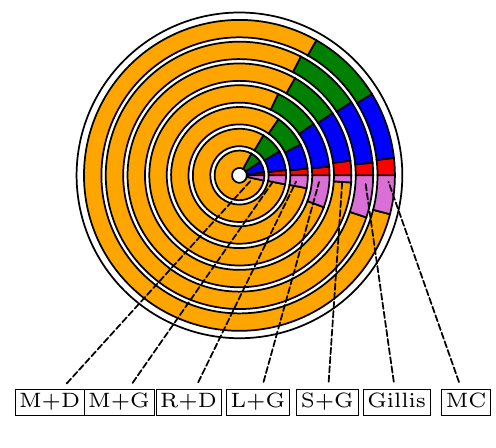}
    \label{fig:s_env_pie_compute}
    }
    \subfigure[Network Constrained]{
    \includegraphics[width=.236\textwidth]{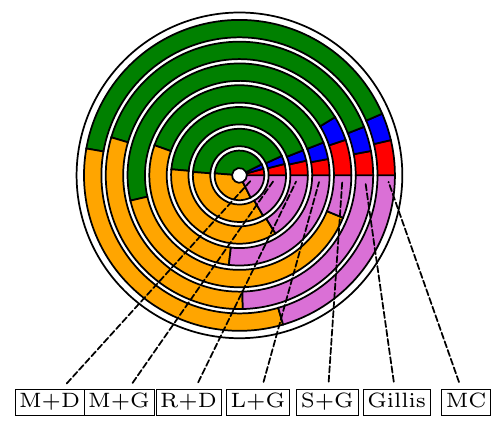}
    \label{fig:s_env_pie_network}
    }
    \subfigure[Memory Constrained]{
    \includegraphics[width=.236\textwidth]{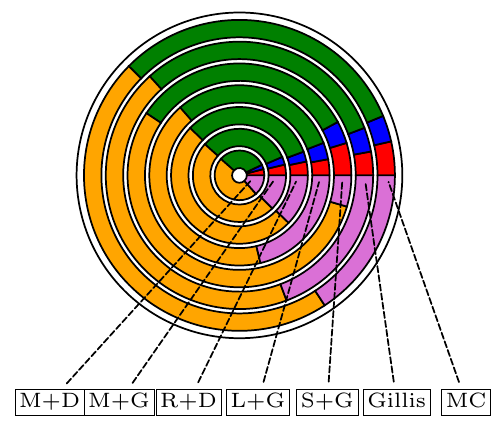}
    \label{fig:s_env_pie_memory}
    }
    \caption{Distribution of response time in Normal and Compute, Network and Memory constrained environments.}
    \label{fig:s_env_pie}
\end{figure*}
\begin{figure*}[t]
    \centering
    \includegraphics[width=.5\textwidth]{images/legend_app2.pdf}\\
    \subfigure[Normal Setup]{
    \includegraphics[width=.236\textwidth]{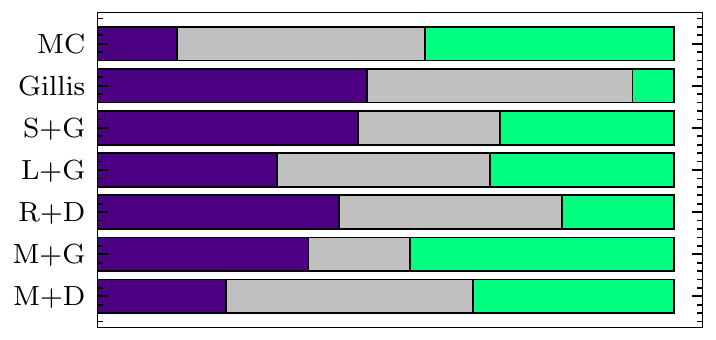}
    \label{fig:s_env_pie_normal_sla}
    }
    \subfigure[Compute Constrained]{
    \includegraphics[width=.236\textwidth]{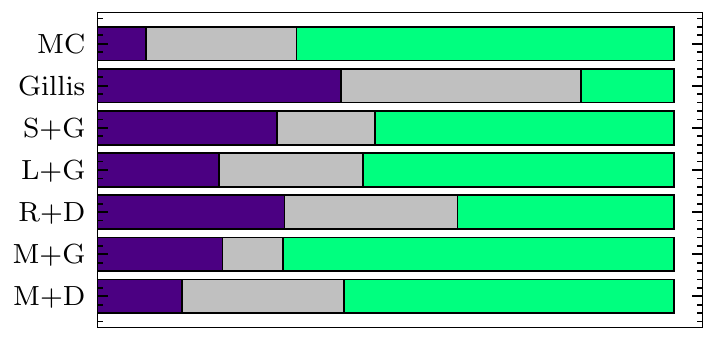}
    \label{fig:s_env_pie_compute_sla}
    }
    \subfigure[Network Constrained]{
    \includegraphics[width=.236\textwidth]{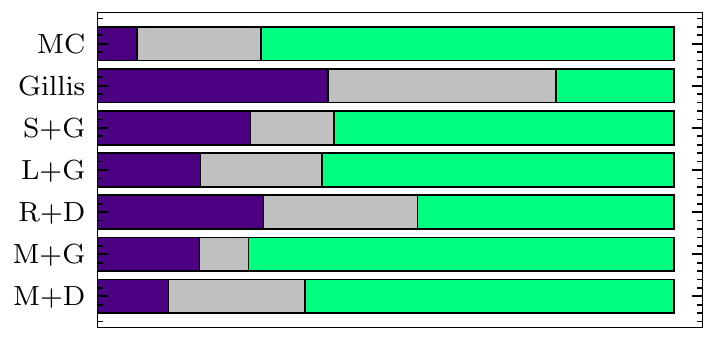}
    \label{fig:s_env_pie_network_sla}
    }
    \subfigure[Memory Constrained]{
    \includegraphics[width=.236\textwidth]{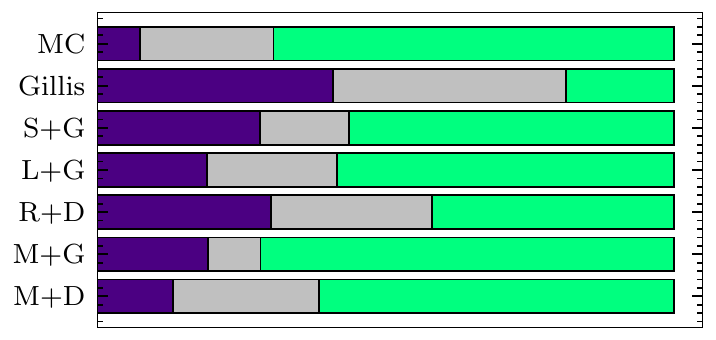}
    \label{fig:s_env_pie_memory_sla}
    }
    \caption{Distribution of SLA violations in Normal and Compute, Network and Memory constrained environments.}
    \label{fig:s_env_pie_sla}
\end{figure*}

\textbf{Results:}
Figure~\ref{fig:s_env} shows the performance metrics for the three constrained environments for all models. The primary $y$-axis gives the value of the metric in the new setting and the secondary $y$-axis gives us the ratio with respect to metric value in the original setup. Figure~\ref{fig:s_env_pie} shows the distribution of the response time to the average time taken for scheduling, waiting, data transfer, task execution and container migration. Figure~\ref{fig:s_env_pie_sla} shows the distribution of the SLA violations for each application. Similar to what we observed earlier, for reward free models (R+D, L+G and S+G), the average accuracy is nearly unchanged, whereas for M+D, M+G, MC and Gillis, the accuracy drop is up to $40\%$. This drop is more pronounced in network and compute constrained environments as compared with the memory constrained one. This is because of the increasing average response times compared to the original setup due to lower computational resource availability, higher migration and data response times or low memory resources. This leads to higher resource contention in these devices, again for which the MAB model tends to take a higher ratio of semantic-split decisions to avoid too many SLA violations. This leads to the decline in accuracy of the M+D, M+G and baseline models in the constrained environments, whereas the average accuracy for reward free models (R+D, L+G, S+G) remains nearly unchanged.  However, the response times and fraction of SLA violations are much higher for the reward free models. The relative increase in the violation rates are much lower for the M+D and M+G policies, thanks to the intelligent adaptability of the MAB model to the changing response times in the new setups. The increase in the response times in the case of computational constraints is predominantly due to the increase in the execution times due to low compute resources available (Figure~\ref{fig:s_env_pie_compute}). For network constrained setup, the increase in response time is majorly due to the increase in data transfer and container migration times (due to the lower bandwidth availability, see Figure~\ref{fig:s_env_pie_network}). In memory constrained setup, low memory availability causes the edge devices to use the disk-swap space increasing both execution and data transfer times (Figure~\ref{fig:s_env_pie_memory}). These constraints have the highest impact on the resource hungry CIFAR100 workloads which leads to an increase in its SLA violation rate compared to other application types (Figure~\ref{fig:s_env_pie_sla}).

The average rewards for all models are lower than the original setup (Figure~\ref{fig:s_env_reward}). The relative rewards for MAB based models are the highest ($0.84-0.95$) compared to the baseline models ($0.63-0.0.86$). The reward drop is the least for the SplitPlace model $11.61\%$. The reward drops are higher for the reward-free models ($14.05\%-22.91$), due to their significantly higher SLA violation rate, even without any accuracy change. For all the three cases, the SplitPlace model has the highest average reward ($0.77-0.90$), whereas the Gillis and MC models have rewards in the range $0.58-0.78$ and $0.53-0.71$ respectively. The increase in the SLA violation rates are also the lowest for the SplitPlace model ($9.24\%-22.01\%$) compared to the baselines with $34.73\%-103.11\%$. Similarly, the decrease in the average accuracy for the SplitPlace model is $9.31\%$, which for the baseline models is $16.98\%-32.13\%$. The average increase in the response time in the constrained setup for the SplitPlace model is $14.56\%$, whereas for the baselines it is $12.31\%$. 

\begin{figure*}[t]
    \centering
    \includegraphics[width=.8\textwidth]{images/legend_supp.pdf}\\
    \subfigure[Average Accuracy]{
    \includegraphics[width=.236\textwidth]{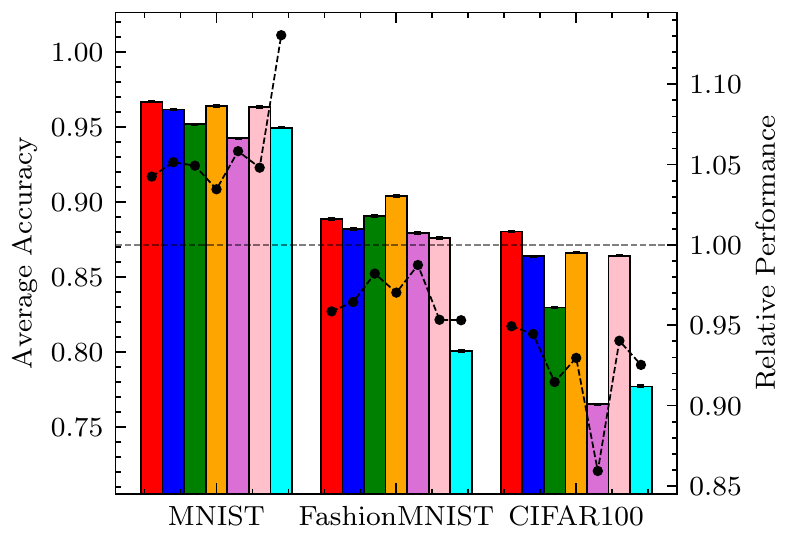}
    \label{fig:s_work_a}
    }
    \subfigure[Average Response Time]{
    \includegraphics[width=.236\textwidth]{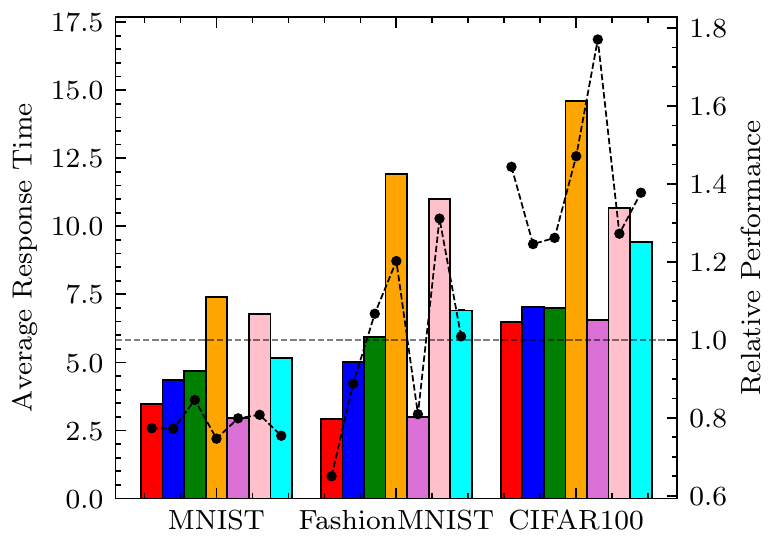}
    \label{fig:s_work_r}
    }
    \subfigure[Fraction of SLA Violations]{
    \includegraphics[width=.236\textwidth]{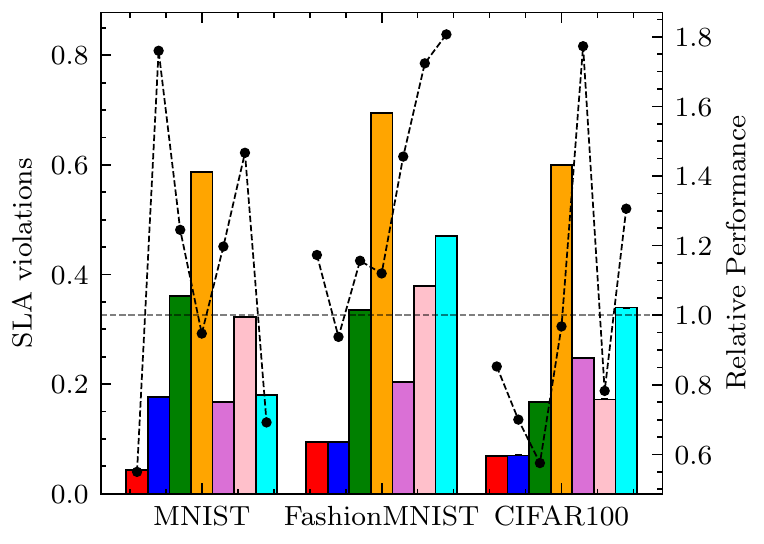}
    \label{fig:s_work_sla}
    }
    \subfigure[Average Reward]{
    \includegraphics[width=.236\textwidth]{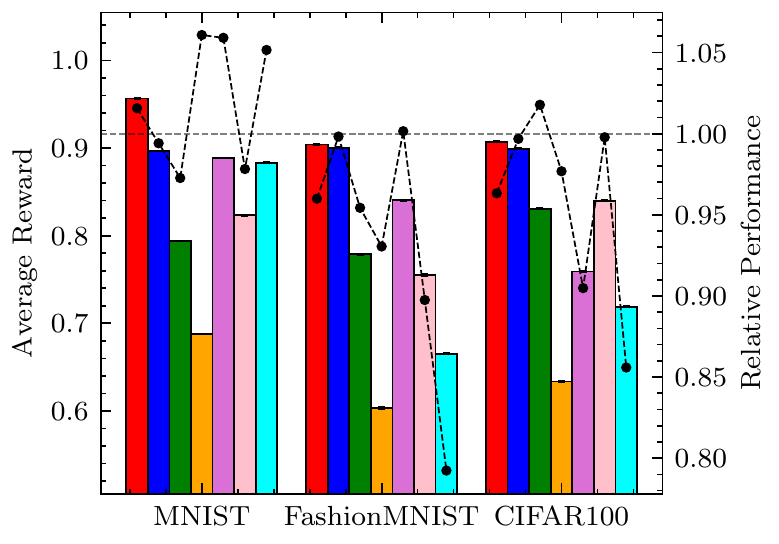}
    \label{fig:s_work_reward}
    }
    \caption{Comparison of SplitPlace against baselines and ablated models in settings with only MNIST, FashionMNIST and CIFAR100 workloads.}
    \label{fig:s_work}
\end{figure*}
\begin{figure*}[t]
    \centering
    \includegraphics[width=.9\textwidth]{images/legend_times.pdf}\\
    \subfigure[Normal Workload]{
    \includegraphics[width=.236\textwidth]{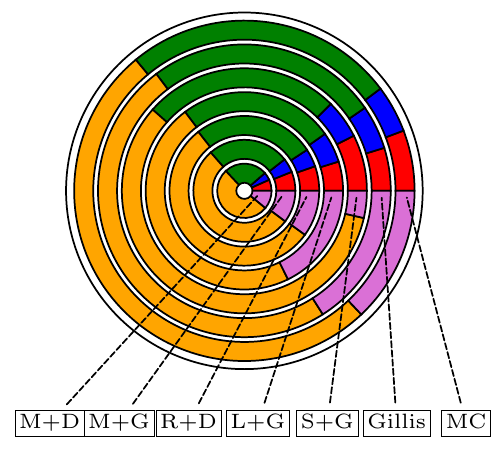}
    \label{fig:s_work_pie_normal}
    }
    \subfigure[MNIST only]{
    \includegraphics[width=.236\textwidth]{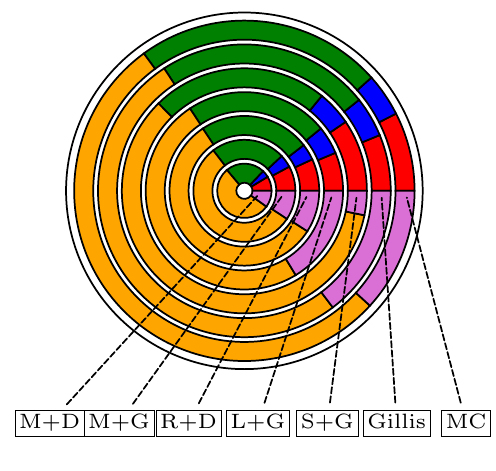}
    \label{fig:s_work_pie_mnist}
    }
    \subfigure[FashionMNIST only]{
    \includegraphics[width=.236\textwidth]{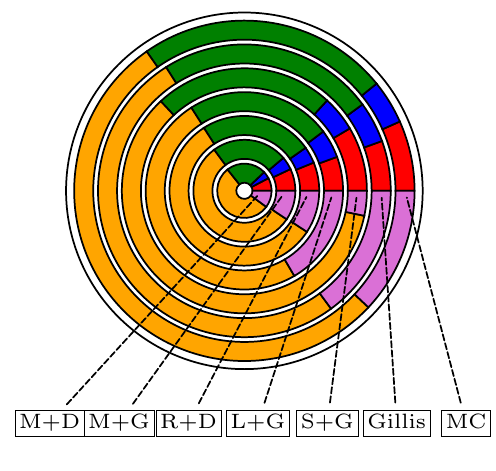}
    \label{fig:s_work_pie_fashionmnist}
    }
    \subfigure[CIFAR100 only]{
    \includegraphics[width=.236\textwidth]{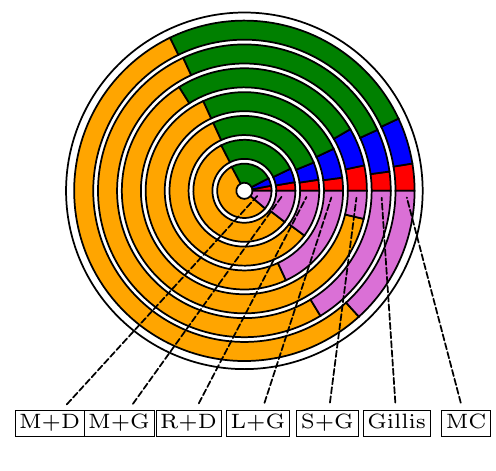}
    \label{fig:s_work_pie_cifar}
    }
    \caption{Distribution of response time in Normal and MNIST only, FashionMNIST only, CIFAR100 only workloads.}
    \label{fig:s_work_pie}
\end{figure*}

\subsection{Constraining Workloads}
\label{app:workload}
We now perform the same experiments, but now changing the workloads to MNIST only, FashionMNIST only and CIFAR100 only applications in lieu of sampling workloads uniformly at random from one of these. This helps us illustrate the robustness of the SplitPlace model against diverse application domains.

\textbf{Results:} Figure~\ref{fig:s_work} shows the performance parameters for all models in the three workload settings.  The graphs show contrasting workload specific trends. For instance, the accuracy of the MNIST application is higher and response time lower than the other two application types. Thus, for all models, the average accuracy is higher compared to the original workload and response time lower. However, due to lower resource requirements of the MNIST workload, the MC model seldom takes the decision to use compressed models, giving a high accuracy uplift compared to the original setup. Figure~\ref{fig:s_work_pie} shows the distribution of the response times of tasks for each workload setting. There are no drastic changes in the time distribution compared to the original workloads, the only difference being the different execution and data transfer times for various workloads. For smaller workloads like MNIST, the execution time and data transfer times are lower than larger workloads like CIFAR100.  Considering the average response time and SLA violation rates, the SplitPlace model gives a lower average violations in all three cases. This is due to the limited workload complexity and simpler decision making for the MAB model. 

Even with these constrained workloads, the SplitPlace model has the highest average accuracy with the lowest average response times and SLA violation rates. Overall, we see that the average reward of the SplitPlace model ($89.83-95.36$) is highest among all methods. The baseline models, Gillis and MC have rewards in the range $75.54-83.98$ and $66.57-88.31$ respectively.

\begin{figure}[t]
    \centering
    \includegraphics[width=.8\linewidth]{images/legend_app2.pdf}\\
    \subfigure[Response Time]{
    \includegraphics[width=.4\linewidth]{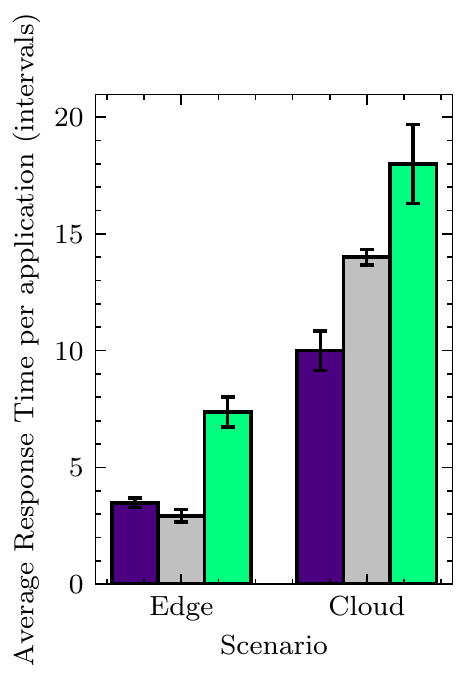}
    \label{fig:cloudai_rt}
    }
    \subfigure[SLA Violation Rate]{
    \includegraphics[width=.4\linewidth]{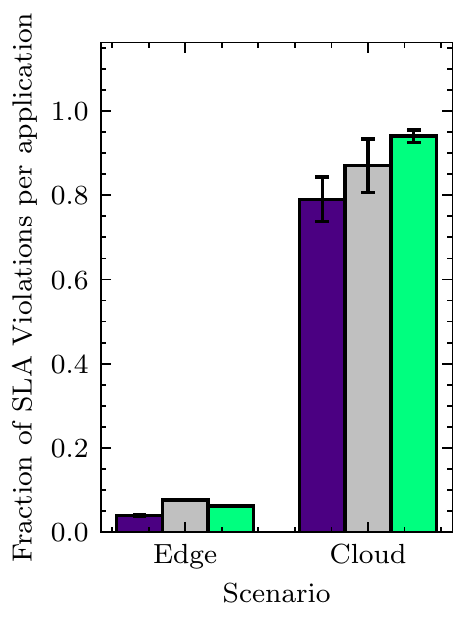}
    \label{fig:cloudai_sla}
    }
    \caption{Comparing the response time and SLA violation rates for SplitPlace for edge and cloud setups.}
    \label{fig:cloudai}
\end{figure}

\subsection{Comparing Edge with Cloud}
\label{app:cloudai}

We also perform experiments to compare the performance of cloud versus edge platforms. As motivated in Section~\ref{sec:introduction}, for latency-critical applications, it is crucial to resort to edge devices that are in close proximity with the users. Relying on cloud nodes, which may be at multi-hop distance from the users lead to higher latency and subsequently the average response times. Thus, for latency critical applications, utilizing memory abundant cloud nodes, with typically 32-64 GB of memory available, is not feasible. This makes the problem challenging as now we need to ensure that we can run large-scale DNNs on edge, without any cloud backend whatsoever. Thus, in our formulation and experiments, we consider an environment with only memory-constrained edge nodes with 4-8 GB RAM and all in proximity with the users. To empirically verify this, we consider a "Cloud" setup wherein the broker nodes remains in the Azure UK-South datacenter, whereas the worker nodes are now initialized in the East-US datacenter. \blue{For a fair comparison, we consider a setup where no splitting is required and end-to-end models are used in the cloud nodes.} The response time and SLA violation rates are shown in Figure~\ref{fig:cloudai}. The figure demonstrates that in a Cloud-AI setup, the same worker nodes are unable to conform to the strict deadlines when compared to the distributed Edge-AI setup. \blue{In terms of the initial one-time communication for transferring the containers, SplitPlace method on edge takes 30 seconds, whereas the cloud-only approach takes 72 seconds. The single-hop communication topology among the broker-worker nodes foregoes the need for sending the container images via the Internet.}

\begin{figure}[t]
    \centering
    \includegraphics[width=.8\linewidth]{images/legend_app2.pdf}\\
    \subfigure[Splitting Decision]{
    \includegraphics[width=.4\linewidth]{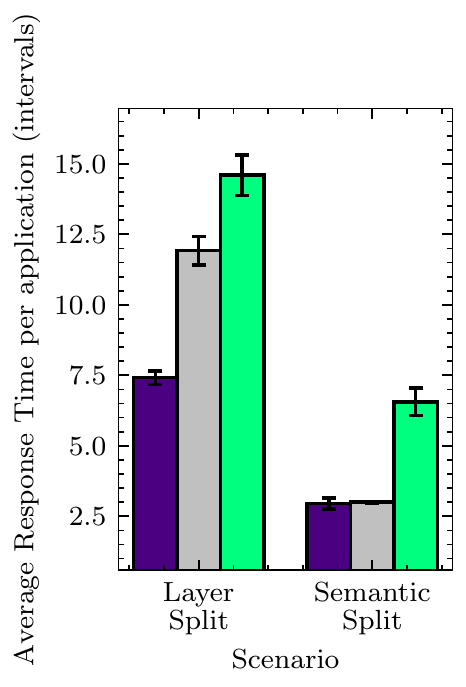}
    \label{fig:svp_split}
    }
    \subfigure[Placement Decision]{
    \includegraphics[width=.4\linewidth]{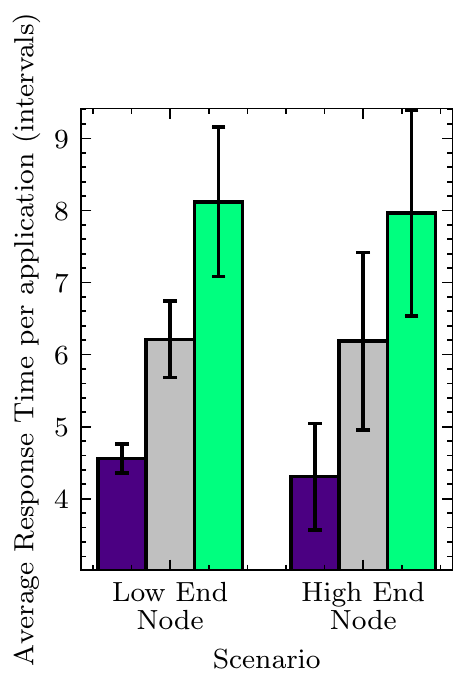}
    \label{fig:svp_place}
    }
    \caption{Comparing the response time across splitting and placement decisions.}
    \label{fig:svp}
\end{figure}

\subsection{Comparing the Impact of Splitting and Placement Decisions on Response Time}
\label{app:splitplace}

The decomposition of decision making in SplitPlace is based on the hypothesis that the MAB models do not need split placement decisions to decide the splitting decisions. The major contributing factor for response time of a task is the split decision type a shown in Figure~\ref{fig:layer_sem}. As the figure demonstrates, the deviation in the response time is much higher when we compare layer vs semantic. The deviation is low when comparing across placement decisions. This is demonstrated by Figure~\ref{fig:svp}. Even though we consider a heterogeneous edge computing environment, with edge devices having different resource capacities, the difference in resources is not sufficient to give significant deviation in terms of response time as per the placement decisions. %\blue{Thus, to avoid the curse of dimensionality adversely impacting the MAB models, we only consider splitting decisions (\textit{viz} a binary decision compared to a more choices for placement) and decompose the into the \underline{split}ting using MABs and \underline{place}ment using neural optimization; hence, the name SplitPlace.}

\end{document}